\documentclass[a4paper,fleqn,usenatbib]{mnras}

\usepackage{newtxtext,newtxmath}
\usepackage[T1]{fontenc}
\usepackage{ae,aecompl}

\usepackage[caption=false]{subfig}
\usepackage{graphicx}	% Including figure files
\usepackage{amsmath}	% Advanced maths commands
\usepackage{amssymb}	% Extra maths symbols
\usepackage[mathscr]{euscript}
\newcommand{\D}{\boldsymbol{\mathsf{D}}}	% Bold diffusivity

\newcommand{\brvs}{Br\"unt--V\"ais\"al\"a}

\title[Enhanced Mixing in Massive Rotating Stars]{Enhanced Rotational Mixing in the Radiative Zones of Massive Stars}

\author[Adam S. Jermyn et al.]{
Adam S. Jermyn,$^{1}$\thanks{E-mail: adamjermyn@gmail.com}
Christopher A. Tout$^{1}$
and
Shashikumar M. Chitre$^{1,2}$
\\
$^{1}$Institute of Astronomy, University of Cambridge, Madingley Rd, Cambridge CB3 0HA, UK\\
$^{2}$Centre for Basic Sciences, University of Mumbai, India\\
}

\date{Accepted XXX. Received YYY; in original form ZZZ}

\pubyear{2017}

\begin{document}
\label{firstpage}
\pagerange{\pageref{firstpage}--\pageref{lastpage}}
\maketitle

\begin{abstract}
Convection in the cores of massive stars becomes anisotropic when they rotate.
This anisotropy leads to a misalignment of the thermal gradient and the thermal flux, which in turn results in baroclinicity and circulation currents in the upper radiative zone.
We show that this induces a much stronger meridional flow in the radiative zone than previously thought.
This drives significantly enhanced mixing, though this mixing does not necessarily reach the surface.
The extra mixing takes on a similar form to convective overshooting, and is relatively insensitive to the rotation rate above a threshold, and may help explain the large overshoot distances inferred from observations.
This has significant consequences for the evolution of these stars by enhancing core-envelope mixing.
\end{abstract}


%%%%%%%%%%%%%%%%%%%%%%%%%%%%%%%%%%%%%%%%%%%%%%%%%%

%%%%%%%%%%%%%%%%% BODY OF PAPER %%%%%%%%%%%%%%%%%%

\section{Introduction}

The problem of meridional mixing in the radiative zones of rotating stars has been studied for nearly a century \citep{1929MNRAS..90...54E, 1998A&A...334.1000M}.
The understanding that has emerged is based on the famed von Zeipel theorem~\citep{Zeipel14071924}, which states that in a radiative zone there is no way to satisfy the equations of hydrostatic and thermal equilibrium without allowing the pressure and temperature to vary separately.
Hence in rotating stars thermodynamic quantities are not solely determined by the pressure; they exhibit baroclinicity.
This baroclinicity drives a circulation current that mixes material in the meridional plane.

The scale of this current is proportional to the heat flux, to the reciprocal of the entropy gradient and to $\Omega^2 R/g$, where $\Omega$ is the angular velocity and $g$ and $R$ are the typical gravity and cylindrical radial coordinate in the radiative zone.
The interpretation of this effect, either as due to transient damping or a genuinely driven flow, has led to some controversy \citep{1981GApFD..17..215B} but it is generally agreed that the mixing itself occurs~\citep{1982PASJ...34..257O}.

Underlying the standard analysis of this mixing is the assumption that baroclinicity is generated locally by the requirements of thermal and hydrostatic equilibrium.
A key result of this work is that this fails in stars with central convection zones\footnote{These are those with masses above $1.2 \mathrm{M}_{\sun}$}.
In such stars rotation distorts convective motions, which set a baroclinic boundary condition at the base of the radiative zone.
This boundary condition is often more significant than the locally-generated baroclinicity and so results in enhanced circulation.
This scenario has not been adequately studied and constitutes a novel driver for meridional flows in the radiative zones of massive rotating stars.

A related scenario that has been studied extensively is that of circulation driven by mechanical pumping in the convection zone.
For instance~\citet{2008ApJ...674..498G} and~\citet{2009ApJ...704....1G} examined the transmission of mechanical forcing across the tachocline and~\citet{2010ApJ...719..313G} extended this to stars with multiple convection zones, where there may be interesting interactions between the zones mediated by the intervening radiative zone.
A key insight of these studies which also holds for our scenario is that the circulation cannot be transmitted and propagated without significant mechanical stresses in the radiative zone.
Such stresses have also been considered as drivers of circulation~\citep{1992A&A...265..115Z, 1998A&A...334.1000M} and so must be taken into account in any analysis of these phenomena.

The possibility of enhanced mixing in these systems is particularly relevant in light of the recent discovery of gravitational waves from merging binary black holes \citep{PhysRevLett.116.061102}.
This has led to increased interest in formation mechanisms for massive stellar cores.
Many such mechanisms are quite sensitive to the magnitude of rotational mixing, particularly in massive stars \citep{refId0}, and so understanding this phenomenon in more depth is extremely important.

\begin{figure}
	\includegraphics[width=0.47\textwidth]{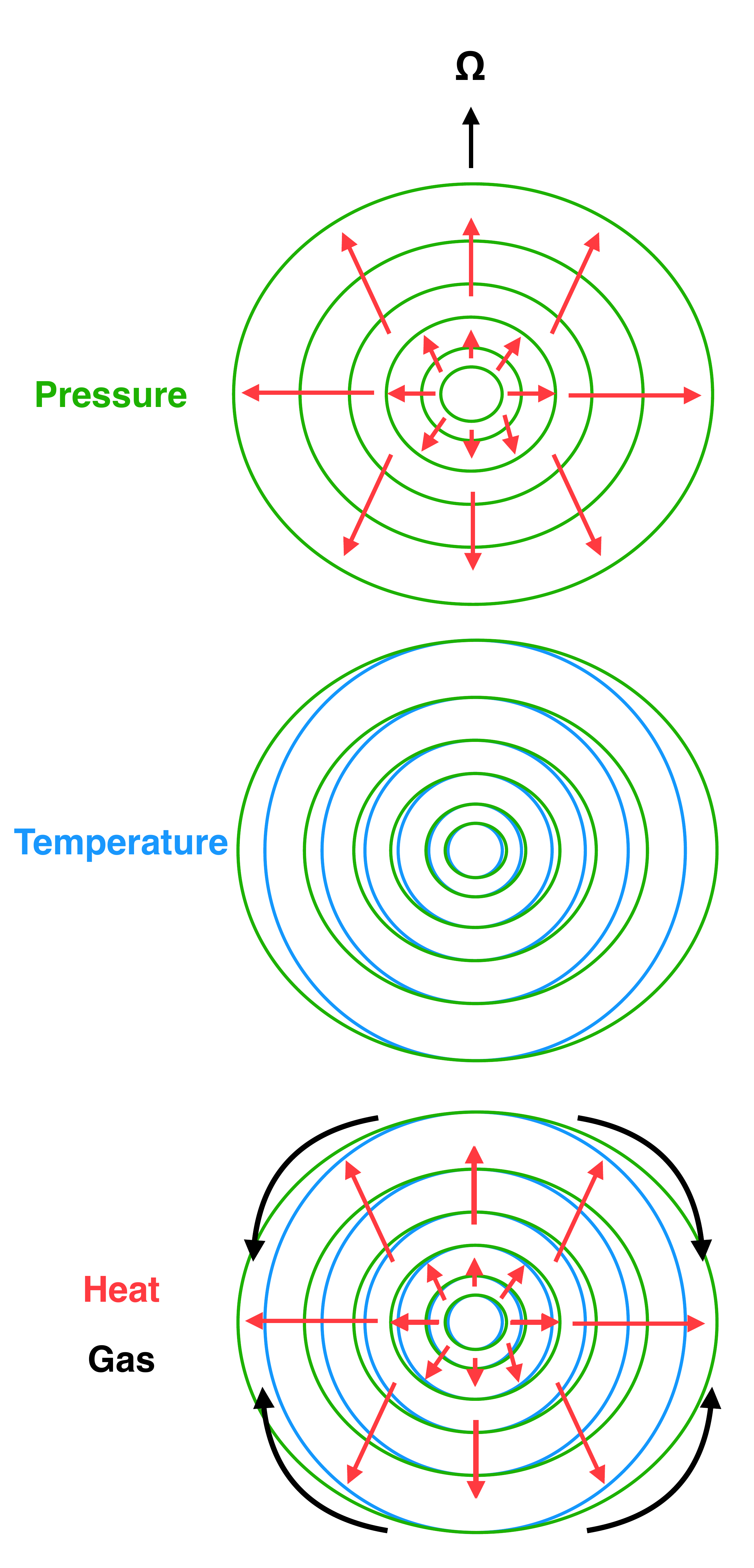}
	\label{fig:es}
	\caption{Centrifugal effects distort isobars. Variation in the magnitude of the pressure gradient along isobars leads to a flux anisotropy (top panel). This pushes the system out of thermal equilibrium, and a new equilibrium is established with baroclinicity (mismatched entropy and pressure surfaces, middle panel) and circulation currents (bottom panel). The length of the arrows is not meaningful. Likewise the geometry of the gas circulation is only meant schematically.}
\end{figure}

In Section~\ref{sec:origins} we review the derivation of the meridional velocity by~\citet{1929MNRAS..90...54E},~\citet{1950MNRAS.110..548S} and~\citet{Zeipel14071924}.
Briefly, the centrifugal effect distorts isobars, causing a flux anisotropy which violates thermal equilibrium (Fig.~\ref{fig:es}, top panel).
Thermal equilibrium is then restored by inducing baroclinicity (Fig.~\ref{fig:es}, middle panel) and hence introducing a circulation current (Fig.~\ref{fig:es}, bottom panel).

\begin{figure}
	\includegraphics[width=0.47\textwidth]{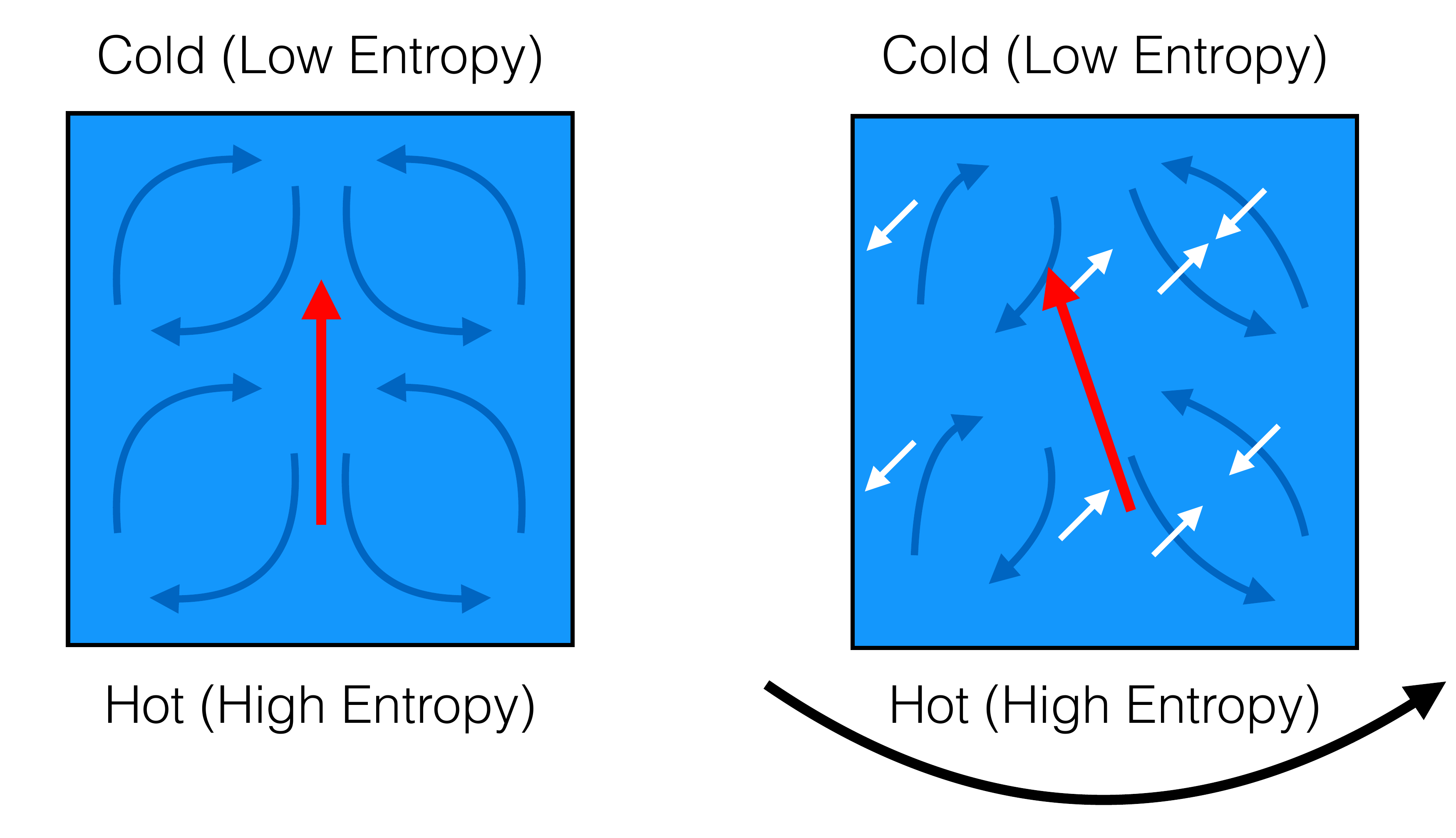}
	\caption{Non-rotating convection (left, blue arrows) transports heat along the entropy gradient. Rotating convection (right, blue arrows) is distorted by the Coriolis force (white arrows) and so transports heat at an angle relative to the entropy gradient.}
	\label{fig:coriolis}
\end{figure}

In Section~\ref{sec:turb} we argue that turbulent anisotropy in the convection zone makes the zone baroclinic.
This arises because the Coriolis effect couples different components of the velocity\footnote{We also estimate the effect of centrifugally--induced baroclinicity and find it to be quite a bit smaller.} and so produces heat transport at an angle to the entropy gradient, as shown in Fig.~\ref{fig:coriolis}.
As in the case of radiative zones this results in a circulation current.
Notably this is significantly enhanced relative to what would occur in a radiative zone because the radial entropy gradient is much smaller in convection zones.

A further consequence of turbulent anisotropy is that the heat flux in the zone is highly aspherical\footnote{This is related to the baroclinicity because the convective flux depends strongly on the entropy gradient.}.
This is important in massive stars with convective cores because this flux perturbation produces baroclinicity at the base of the radiative zone.
In Section~\ref{sec:massive} we argue that this baroclinic boundary condition serves to drive additional mixing in the radiative zone.
This mixing is often orders of magnitude greater than what the Eddington--Sweet calculation alone provides.
We then comment on the prospects for similar mechanisms in stars of lower mass.

In Section~\ref{sec:chemistry} we argue that baroclinic mixing mechanism is not fundamentally altered by chemical composition gradients.
In particular we show that such gradients play the same role in this mechanism as they do in the Eddington--Sweet circulation and hence only produce corrections of order unity.

Finally in Section~\ref{sec:overshoot} we argue that these circulation currents are actually the source of the extra mixing normally attributed to convective overshooting.
We then discuss the physical interpretation of the circulation in light of historical difficulties with the Eddington--Sweet circulation.

\section{Origins of Meridional Flow}
\label{sec:origins}

The meridional velocity arises from the impossibility of thermal and hydrostatic equilibrium when the pressure gradient is misaligned with the thermal flux.
In the case considered by~\citet{1929MNRAS..90...54E} and von Zeipel~\citep{Zeipel14071924} this arises because of a bending of the isobars owing to $\mathcal{O}(\Omega^2)$ centrifugal effects.
We now review this derivation.

To begin note that, in the absence of differential rotation, there exists an effective potential satisfying
\begin{equation}
\nabla^2 \Phi = 4\pi G\rho - 2\Omega^2,
\label{eq:poisson}
\end{equation}
where $\rho$ is the density and $G$ is Newton's gravitational constant.
With hydrostatic equilibrium we also have
\begin{equation}
\nabla P = -\rho \nabla \Phi
	\label{eq:hydroeq}
\end{equation}
This means that
\begin{align}
	\nabla\times\left(\frac{1}{\rho}\nabla P\right) = -\nabla\times\nabla\Phi = 0
\end{align}
and so
\begin{align}
	\nabla \rho \times \nabla P = 0,
	\label{eq:baro}
\end{align}
hence isobars and isochors coincide.
From equation~\eqref{eq:hydroeq} it follows that these coincide with surfaces of constant $\Phi$ and so $P$ and $\rho$ may be written solely as functions of $\Phi$.

The energy balance in stars is given by
\begin{equation}
	c_p \rho T \frac{\partial s}{\partial t} = \frac{\partial F_i}{\partial x_i} + \rho \varepsilon - \rho c_p T u_i\frac{\partial s}{\partial x_i},
	\label{eq:thermalbalance}
\end{equation}
where $c_p$ is the specific heat capacity at constant pressure, $\boldsymbol{F}$ is the diffusive flux, $\varepsilon$ is the specific rate of energy generation from nuclear processes, $s$ is the dimensionless entropy, $\rho$ is the density, $T$ is the temperature, $u_i$ is the velocity and summation is implicit over repeated indices.
In the case of radiation we may write the diffusive flux as
\begin{equation}
	F_i = -\rho c_p \chi \frac{\partial T}{\partial x_i},
	\label{eq:diffflux}
\end{equation}
where $\chi$ is the thermal diffusivity.
In steady state $\partial_t s$ vanishes and with no nuclear burning we may set $\varepsilon=0$ so that equation~\eqref{eq:thermalbalance} becomes
\begin{equation}
	\frac{\partial F_i}{\partial x_i} - \rho c_p T u_i\frac{\partial s}{\partial x_i} = 0.
	\label{eq:windeq}
\end{equation}
Now suppose that there is no meridional flow, such that $u=0$.
We have then
\begin{equation}
		\frac{\partial F_i}{\partial x_i} = 0.
	\label{eq:nowindeq0}
\end{equation}
So from equation~\eqref{eq:diffflux}
\begin{equation}
	\frac{\partial}{\partial x_i}\left(\rho c_p \chi \frac{\partial T}{\partial x_i}\right) = 0.
	\label{eq:nowindeq}
\end{equation}
All these quantities are thermodynamic variables which may be written as functions of $P$ and $\rho$ using the equation of state.
Because $P$ and $\rho$ are functions only of $\Phi$ equation~\eqref{eq:nowindeq} may be written as
\begin{equation}
	\frac{\partial}{\partial x_i}\left(q(\Phi) \frac{\partial \Phi}{\partial x_i}\right) = 0,
	\label{eq:nowindeq1}
\end{equation}
where $q$ is a function only of potential.
Expanding the divergence yields
\begin{equation}
	\frac{dq}{d\Phi} \left|\nabla \Phi\right|^2 + q(\Phi) \nabla^2 \Phi = 0
	\label{eq:nowindeq2}
\end{equation}
and hence
\begin{equation}
	\left|\nabla \Phi\right|^2 = - \frac{q(\Phi) \nabla^2 \Phi}{\mathrm{d}q/\mathrm{d}\Phi}.
	\label{eq:nowindeq3}
\end{equation}
The right-hand side may be evaluated with equation~\eqref{eq:poisson} to find a function only of $\Phi$.
The left-hand side, by contrast, may be written as
\begin{equation}
\left|\nabla \Phi\right|^2 = \left|\boldsymbol{g} - \Omega^2 R \boldsymbol{e}_{r}\right|^2,
\end{equation}
where $R$ is the distance from the rotation axis, $\boldsymbol{e}_{r}$ is the cylindrical radial unit vector and $\boldsymbol{g}$ is the acceleration due to gravity.
This generally varies along isobars because the direction of $\boldsymbol{g}$ changes while that of the centrifugal acceleration does not, and the magnitude of the centrifugal acceleration varies while, to leading order, that of gravity does not.
As a result equation~\eqref{eq:nowindeq3} cannot be satisfied simultaneously with equation~\eqref{eq:baro}.
This is the original Eddington--Sweet argument for radiative zones \citep{Zeipel14071924}.

From this it follows that there must at a minimum be either a meridional flow or baroclinicity or both.
The original Eddington--Sweet argument does not address the possibility of baroclinicity as an alternative and so just concludes that a meridional flow arises with scale set by equation~\eqref{eq:windeq}.
In Appendix~\ref{appen:diffrot} we address this point and demonstrate that in fact both a meridional flow and differential rotation generally arise, with this equation setting the scale of the flow.

\section{Anisotropic Convection}
\label{sec:turb}

We now turn to the convection zone with the aim of showing that convective anisotropy results in baroclinicity.
When viewed on large length-scales convective turbulence acts to diffuse heat\footnote{More formally, the convective diffusivity arises from applying the Renormalization Group to the Navier--Stokes equation. Upon integrating out short-range degrees of freedom we obtain the effective couplings among longer-range degrees of freedom~\citep[See e.g.][]{1986JSCom...1....3Y}. The motion is random and so leads to diffusion rather than advection.}.
The effective diffusivity tensor is of the form
\begin{equation}
	\D_{ij} \approx \langle \delta x_i \delta u_j\rangle.
\end{equation}
Physically this just means that material located at $\delta \boldsymbol{x}$ relative to the centre of an eddy is transported along with velocity $\delta \boldsymbol{u}$.

In the presence of an entropy gradient, this diffusivity gives rise to a heat flux
\begin{equation}
	F_i = - c_\mathrm{p} \rho T \mathsf{D}_{ij} \frac{\partial s}{\partial x_j}.
	\label{eq:diffflux2}
\end{equation}
Crucially, the convective viscosity is not isotropic~\citep{2013MNRAS.431.2200L, gough78,1957ApJ...126..259U}.
This is because the Coriolis acceleration of a fluid parcel owing to its velocity $\boldsymbol{u}$ is
\begin{equation}
	\boldsymbol{a}_\mathrm{c} = 2 \boldsymbol{u}\times \boldsymbol{\Omega},	
\end{equation}
where $\boldsymbol{\Omega}$ is the angular velocity of rotation.
Because this expression contains a cross product, the Coriolis effect generically leads to a coupling between different components of the velocity, and hence between different components of the position and velocity of the eddy.
In rotating systems this means that the heat flux is not aligned with the entropy gradient~\citep{1993A&A...276...96K, 2013MNRAS.431.2200L}, and it is this effect that drives baroclinicity.

\begin{figure}
	\centering
	\includegraphics[width=0.47\textwidth]{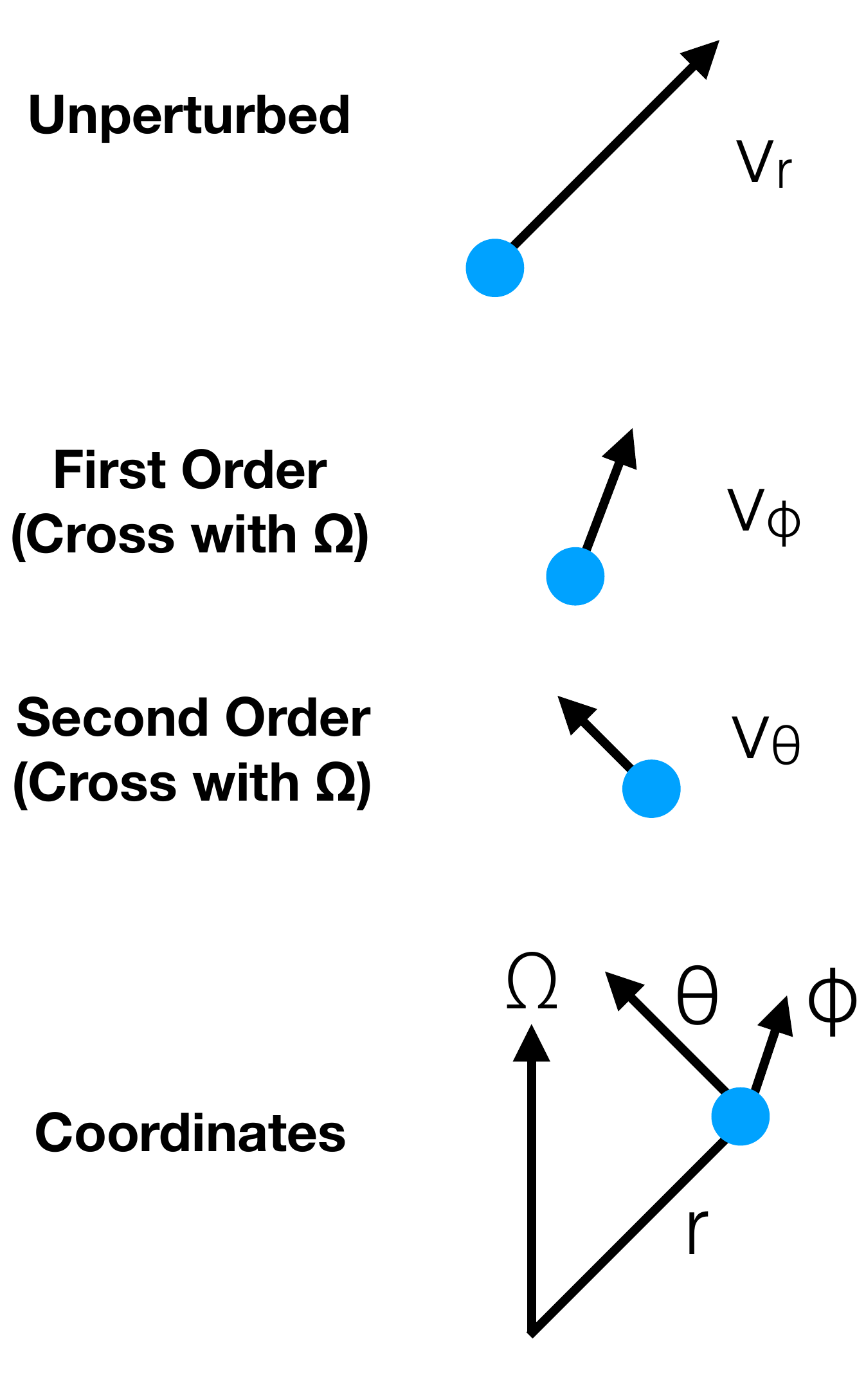}
	\label{fig:coriolis2}
	\caption{The radial motion $\varv_r$ of a convective eddy is shown unperturbed. The Coriolis force acting on this creates an azimuthal component $u_\theta$ at first order. Acting on this the Coriolis force creates a latitudinal component $u_\theta$ at second order. So two applications of the Coriolis effect are required to generate motion along $\boldsymbol{e}_\theta$ from motion along $\boldsymbol{e}_r$.}
\end{figure}

In spherical coordinates the diffusivity tensor may be written as
\begin{equation}
\mathbfss{D} = D_0 \mathbfss{D}' = D_0\left( \begin{array}{ccc}
1 & \epsilon_{\theta r} & \epsilon_{\phi r} \\
\epsilon_{r\theta} & c_1 & \epsilon_{\phi\theta} \\
\epsilon_{r \phi} & \epsilon_{\theta\phi} & c_2 \end{array} \right),
\label{eq:nuexp}
\end{equation}
where $D_0$ is a scalar function~\citep{2012ISRAA2012E...2G}.
When $\Omega=0$ this tensor is diagonal because of spherical symmetry.
The terms on the diagonal do not vanish when $\Omega=0$, so we let $c_1$ and $c_2$ be constants of order unity.
Perturbing away from this limit we argue that all components involving the azimuthal direction are linear in $\Omega$ because a single application of the Coriolis effect suffices to correlate meridional motion with azimuthal motion.
By contrast $\epsilon_{r\theta}$ and $\epsilon_{\theta r}$ are quadratic in $\Omega$ at leading order because it takes two applications of this effect to correlate one component of meridional motion with another, as shown in Fig.~\ref{fig:coriolis2}.
This form is in agreement with the works of~\citet{2013IAUS..294..399K} as well as those of~\citet{jermyn} and~\citet{2013MNRAS.431.2200L}, each of which have been verified with data from 3D simulations of rotating convection.

To see how this drives baroclinicity we first calculate the corrections that the convective anisotropy introduces in the equation of thermal equilibrium.
For this we assume that there is no meridional flow in the convection zone.
We then introduce a meridional flow in the convection zone and demonstrate that it does not change the overall magnitude of the baroclinicity.

Inserting equations~\eqref{eq:nuexp} and~\eqref{eq:diffflux2} into equation~\eqref{eq:nowindeq0} we find
\begin{equation}
	\left(\nabla + \nabla \ln (\rho T D_0)\right) \cdot \left(\D' \cdot \nabla s\right)  = 0.
	\label{eq:thermalbalance2}
\end{equation}
Taking the system to be axisymmetric, we may write the function $D_0$ as
\begin{equation}
	D_0 = D_{0,0}(P,T) + \Omega^2D_{0,1}(P,T,\theta) + \mathcal{O}(\Omega^3).	
\end{equation}
The dependence of $D_{0,1}$ on $\theta$ and the independence of $D_{0,0}$ on the same reflects the symmetry of the problem: only contributions coupling to the rotation may depend on latitude because in the absence of rotation latitude is an arbitrary coordinate.
In addition, we write
\begin{equation}
	\nabla s = \boldsymbol{e}_r s_r + \boldsymbol{e}_{\theta} s_\theta,	
\end{equation}
where $s_\theta$ is $\mathcal{O}(\Omega^2)$, and
\begin{equation}
	\nabla \ln (\rho T D_0) = \boldsymbol{e}_r t_r + \boldsymbol{e}_{\theta} t_\theta,	
\end{equation}
where $t_\theta$ is $\mathcal{O}(\Omega^2)$.
In both cases we have used axisymmetry, so that $\partial/\partial_\phi \equiv 0$.
Inserting these into equation \eqref{eq:thermalbalance2} gives
\begin{align}
	0=&\left(\nabla + t_r \boldsymbol{e}_r + t_\theta \boldsymbol{e}_{\theta}\right)\cdot\left[\boldsymbol{e}_r\left(s_r + \epsilon_{\theta r}s_\theta\right) + \boldsymbol{e}_{\theta}\left(s_r \epsilon_{r\theta} + s_\theta c_1\right)\right].
	\label{eq:presimp}
\end{align}
Expanding equation~\eqref{eq:presimp} then gives us
\begin{align}
	0 = &\frac{1}{r^2}\frac{\partial}{\partial r}\left(r^2\left(s_r + \epsilon_{\theta r}s_\theta\right)\right) + \frac{1}{r\sin\theta}\frac{\partial}{\partial \theta}\left(\sin\theta\left(s_r \epsilon_{r\theta} + s_\theta c_1\right)\right)\nonumber\\
		&+ t_r (s_r + \epsilon_{\theta r} s_\theta) + t_\theta (s_r \epsilon_{r\theta} + s_\theta c_1).
\end{align}
We now drop terms which we know to be of order $\Omega^3$ and higher so that
\begin{align}
	0 = &\frac{1}{r^2}\frac{\partial}{\partial r}\left(r^2 s_r\right) + \frac{1}{r\sin\theta}\frac{\partial}{\partial \theta}\left(\sin\theta\left(s_r \epsilon_{r\theta} + s_\theta c_1\right)\right)+ t_r s_r.
\end{align}
As before terms which couple to latitude must be at least second order in $\Omega$.
This includes derivatives in $\theta$ of terms which themselves are non-zero for $\Omega=0$.
This allows us to neglect the term $\epsilon_{r\theta} \partial_\theta s_r$ but does not allow any such simplification for the remaining terms because those which are $\mathcal{O}(\Omega^2)$ do not necessarily become of higher order when differentiated with respect to latitude.
Thus we have
\begin{align}
	0 = &\frac{1}{r^2}\frac{\partial}{\partial r}\left(r^2 s_r\right) + \frac{1}{r}\left(s_r \frac{\partial  \epsilon_{r\theta}}{\partial \theta} + c_1\frac{\partial s_\theta}{\partial \theta}\right)\nonumber\\
	& + \frac{\cot\theta}{r}\left(s_r \epsilon_{r\theta} + s_\theta c_1\right)+ t_r s_r.
\end{align}
Reorganising terms we write
\begin{align}
	0  = &\frac{\partial \epsilon_{r\theta}}{\partial \theta} + \frac{c_1}{s_r}\frac{\partial s_\theta}{\partial \theta} + \cot\theta\left(\epsilon_{r\theta} + \frac{s_\theta c_1}{s_r}\right)+2 + r t_r + \frac{\partial \ln s_r}{\partial \ln r}.
\end{align}

Now $\epsilon_{r\theta}$, $\partial_\theta \epsilon_{r\theta}$, $s_\theta$ and $\partial_\theta s_\theta$ vanish when $\Omega=0$ because these terms can only be non-zero when spherical symmetry is broken.
So
\begin{align}
	0 = &\frac{\partial \epsilon_{r\theta}}{\partial \theta} + \frac{c_1}{s_r}\frac{\partial s_\theta}{\partial \theta}+ \cot\theta\left(\epsilon_{r\theta} + \frac{s_\theta c_1}{s_r}\right) +\Delta \left(r t_r + 2 + \frac{\partial \ln s_r}{\partial \ln r}\right),
	\label{eq:thermalBalance3}
\end{align}
where $\Delta$ refers to the difference between the expression evaluated in the rotating and non-rotating cases.
Hence the term in parentheses is at least of order $\Omega^2$.

Because the fourth term in equation~\eqref{eq:thermalBalance3} does not contain an explicit dependence on $\theta$ it may well vanish in a barotropic setting if the other terms all vanish as well.
However if the other terms do not vanish then either their explicit dependence on $\theta$ must cancel among them, in which case it must be that $s_\theta$ is non-zero, or else they must introduce a $\theta$ dependence into the fourth term.
In either case the system becomes baroclinic, and by the same argument as in the Eddington--Sweet case and Appendix~\ref{appen:diffrot} this comes along with a circulation current of the order of the implied baroclinicity.

We have argued that $\epsilon_{r\theta}$ is $\mathcal{O}(\Omega^2)$.
The only relevant\footnote{The dynamical frequency-scale $\sqrt{g/h}$ is also present but is not relevant because the perturbation acts upon convective eddies which move with frequency-scale $|N|$ and does so with coupling proportional to that scale.} time-scale other than the rotation rate is the \brvs\ frequency $|N|$, so up to a dimensionless factor of order unity we expect
\begin{equation}
	\epsilon_{r\theta} \approx \frac{\Omega^2}{|N|^2}.
	\label{eq:eps1}
\end{equation}
Because $\epsilon_{r\theta}$ vanishes on the equator by symmetry we expect its variation over latitudes to be of order unity.
Hence
\begin{equation}
	\frac{\partial\epsilon_{r\theta}}{\partial\theta} \approx \frac{\Omega^2}{|N|^2},
	\label{eq:eps2}
\end{equation}
which is in good agreement with the results of~\citet{1993A&A...276...96K}.
This may also be seen by noting that these quantities are locally determined and are perturbed to second order, so the magnitude of each is just its characteristic scale (unity in the case of $\epsilon_{r\theta}$, $D_0$ in the case of $\Delta D_0$) multiplied by the dimensionless parameter $\Omega^2/|N|^2$.

The quantities $t_r$, $s_\theta$ and $s_r$ and their derivatives, by contrast, are primarily determined non-locally by the perturbing term $\epsilon_{r\theta}$ throughout the atmosphere.
To find them we must integrate outward from the centre of the star, where they all vanish by symmetry.
We approximate these integrals by the integral of the magnitude of the perturbation from the centre to the point of interest.
The relevant radial length scale is the pressure scale height
\begin{align}
	h \equiv -\frac{\partial\ln P}{\partial r},
\end{align}
so radial derivatives of these quantities are characterised by
\begin{equation}
	\left|\frac{d}{dr}\Delta(r t_r)\right| \approx \frac{d}{dr}\left|\frac{s_\theta}{s_r}\right| \approx \left|\frac{d}{dr}\Delta\frac{\partial \ln s_r}{\partial \ln r}\right| \approx \frac{1}{h}\left(\frac{\Omega}{|N|}\right)^2.
	\label{eq:diffpert}
\end{equation}
Integrating up from the core where $P=P_\mathrm{c}$ then gives
\begin{equation}
	\Delta (r t_r) \approx \frac{s_\theta}{s_r} \approx \Delta \frac{\partial \ln s_r}{\partial \ln r} \approx \alpha,
	\label{eq:pert}
\end{equation}
where
\begin{equation}
	\alpha \equiv \int_{\ln P}^{\ln P_{\mathrm{c}}} \frac{\Omega(\mathscr{P})^2}{|N|(\mathscr{P})^2} \mathrm{d}\ln \mathscr{P}
	\label{eq:alpha0}
\end{equation}
and $\mathscr{P}$ is the pressure.
We use this as a proxy for the radial coordinate because even in highly baroclinic regions the pressure gradient is predominantly radial.
Note that $\alpha$ is related\footnote{See Appendix~\ref{appen:lambda}.} to the small angle $\lambda$ between $\nabla p$ and $\nabla \rho$ by
\begin{align}
	\lambda \approx h |s_r \alpha|.
\end{align}
In Appendix~\ref{appen:centrifugal} we demonstrate that the dominant contribution to this is that owing to turbulent anisotropy.
Hence we may neglect centrifugal effects.

\section{Circulation in the Convection Zone}
\label{sec:convvel}

We now examine the consequences for circulation in the convection zone.
We know that it is not possible for rotating radiative zones to avoid circulation currents because they cannot satisfy the condition of thermal equilibrium without becoming baroclinic and, once baroclinic, a flow is generically driven with magnitude set by the baroclinicity.
The same is true for convection zones and so there is a meridional flow.
Incorporating such a flow\footnote{This is done by starting the derivation with equation~\eqref{eq:windeq} rather than equation~\eqref{eq:nowindeq0} and propagating the effects through to equation~\eqref{eq:thermalBalance3}.} and recalling the assumption of axisymmetry, this equation becomes\footnote{Note that $s_\theta$ is $\mathcal{O}(\Omega^2)$ and $u_\theta$ is at least $\mathcal{O}(\Omega)$ so we have dropped their product.}
\begin{align}
	\frac{r}{D_0} u_r = &\frac{\partial \epsilon_{r\theta}}{\partial \theta} + \frac{c_1}{s_r}\frac{\partial s_\theta}{\partial \theta}+ \cot\theta\left(\epsilon_{r\theta} + \frac{s_\theta c_1}{s_r}\right)\nonumber\\
	& +\Delta \left(r t_r + 2 + \frac{\partial \ln s_r}{\partial \ln r}\right),
	\label{eq:thermalBalance4}
\end{align}
where $u_r$ is the radial component of the flow velocity.
From this we see that the magnitude of the flow is set by the typical magnitude of the remaining terms, so we expect
\begin{equation}
	u_r \approx \frac{D_0}{r} \alpha',
	\label{eq:ur0}
\end{equation}
where $\alpha'$ describes the total perturbation to equation~\eqref{eq:thermalBalance3}.
This is related to $\alpha$ by adding the local perturbations given by equations~\eqref{eq:eps1} and~\eqref{eq:eps2}.
Hence
\begin{equation}
\alpha' \equiv \frac{\Omega^2}{|N|^2} + \alpha.
\label{eq:ap}
\end{equation}
This is similar to the result of~\citet{Roxburgh1991}, except that the denominator of the perturbing parameter here is correctly identified as $|N|^2$ rather than $g/r$.

We now determine the relative scales of the radial and latitudinal flows $u_r$ and $u_\theta$.
In steady state conservation of mass gives $\nabla\cdot(\rho \boldsymbol{u})=0$, so
\begin{equation}
	\frac{r}{\sin\theta}\frac{\partial (u_\theta \sin\theta)}{\partial \theta} + r u_\theta\frac{\partial \ln \rho}{\partial \theta} = - \frac{\partial (r^2 u_r)}{\partial r} - u_r r^2 \frac{\partial \ln \rho}{\partial r}.
	\label{eq:varvt0}
\end{equation}
We may drop the term containing $\partial_\theta \ln \rho$ because it is higher-order in $\Omega$ than $\partial_r \ln \rho$.
So
\begin{equation}
	\frac{r}{\sin\theta}\frac{\partial (u_\theta \sin\theta)}{\partial \theta} = - \frac{\partial (r^2 u_r)}{\partial r} - u_r r^2 \frac{\partial \ln \rho}{\partial r},
	\label{eq:varvt}
\end{equation}
We can evaluate the derivatives approximately by noting that $D_0$, and hence $u_r$, have characteristic scale $h$ as does $\rho$.
This is because all of these owe their spatial dependence primarily to the variation of thermodynamic quantities, which ultimately all vary with the same spatial scale as $P$.
This produces the scaling relation
\begin{equation}
	u_\theta \approx u_r \frac{r}{h},	
	\label{eq:urt}
\end{equation}
and thence
\begin{equation}
	u_\theta\approx \frac{D_0}{h} \alpha'.
	\label{eq:ut0}
\end{equation}

From simple mixing length theory of convection
\begin{align}
	D_0 \approx \frac{1}{3} h u_\mathrm{c},
\end{align}
where $u_\mathrm{c}$ is the convective velocity~\citep{gough78}
Inserting this into equations~\eqref{eq:ur0} and~\eqref{eq:ut0} we find
\begin{align}
	u_r &\approx \frac{1}{3} u_\mathrm{c} \frac{h}{r} \alpha'
	\label{eq:ur}
	\intertext{and}
	u_\theta &\approx \frac{1}{3} u_\mathrm{c} \alpha'.
	\label{eq:ut}
\end{align}

As argued in Appendix~\ref{appen:diffrot}, the actual baroclinicity is reduced by virtue of our including the meridional circulation.
In particular, the bulk of the violation of thermal equilibrium implied by equation~\eqref{eq:diffflux2} is resolved by circulation and not by baroclinicity.
So in what follows when we discuss baroclinicity we are talking about that which would exist in the absence of a meridional circulation\footnote{That is, we are using it as a proxy for the anisotropy of the flux distribution.}.
This is the quantity which sets the magnitude of the circulation.

The flow given by equations~\eqref{eq:ur} and~\eqref{eq:ut} generally acts to damp the flux anisotropy (the no-circulation baroclinicity) because it is driven by this effect.
In Appendix~\ref{appen:damp} we show that this modifies equation~\eqref{eq:alpha0} so that
\begin{equation}
	\alpha = P^y\int_{P}^{P_\mathrm{c}} \frac{\Omega(\mathscr{P})^2}{|N|(\mathscr{P})^2} \frac{\mathrm{d}\mathscr{P}}{\mathscr{P}^{1+y}}.
\end{equation}
Note that we work with $\alpha$ rather than directly with the baroclinicity because it is $\alpha$ which evolves according to equation~\eqref{eq:diffpert}.
The effect of the damping is then just to change how we weight the average of the perturbation rather than to change the fundamental scalings.

As a final consideration, equations~\eqref{eq:eps1} and~\eqref{eq:eps2} do not hold for arbitrarily large $\Omega$.
In particular, convective anisotropy is known to saturate for $\Omega$ large relative to $|N|$~\citep{2013MNRAS.431.2200L,2013IAUS..294..399K}.
So we use the prescription
\begin{align}
	\epsilon = \epsilon_{\mathrm{max}}\min\left(1, \frac{\Omega^2}{|N|^2}\right),
\label{eq:epsmax}
\end{align}
where $\epsilon_{\mathrm{max}} \approx 0.2$~\citep{jermyn}, and hence find that
\begin{equation}
	\alpha \approx P^y \int_{P}^{P_\mathrm{c}} \epsilon_{\mathrm{max}}\min\left(1, \frac{\Omega(\mathscr{P})^2}{|N|(\mathscr{P})^2}\right) \frac{\mathrm{d}\mathscr{P}}{\mathscr{P}^{1+y}}.
	\label{eq:alphaPy}
\end{equation}
Furthermore, in the limit of rapid rotation it is no longer the case that there need be a circulation of order the implied baroclinicity, because the baroclinicity alone suffices to restore thermal equilibrium\footnote{See Appendix~\ref{appen:diffrot}.}.
This does not mean that the flux imbalance is less than equation~\eqref{eq:alphaPy} implies but it does imply that $y$ may be smaller in this limit and that equations~\eqref{eq:ut} and~\eqref{eq:ur} may not hold in rapidly rotating convection zones.

\begin{figure}
\centering
\includegraphics[width=0.45\textwidth]{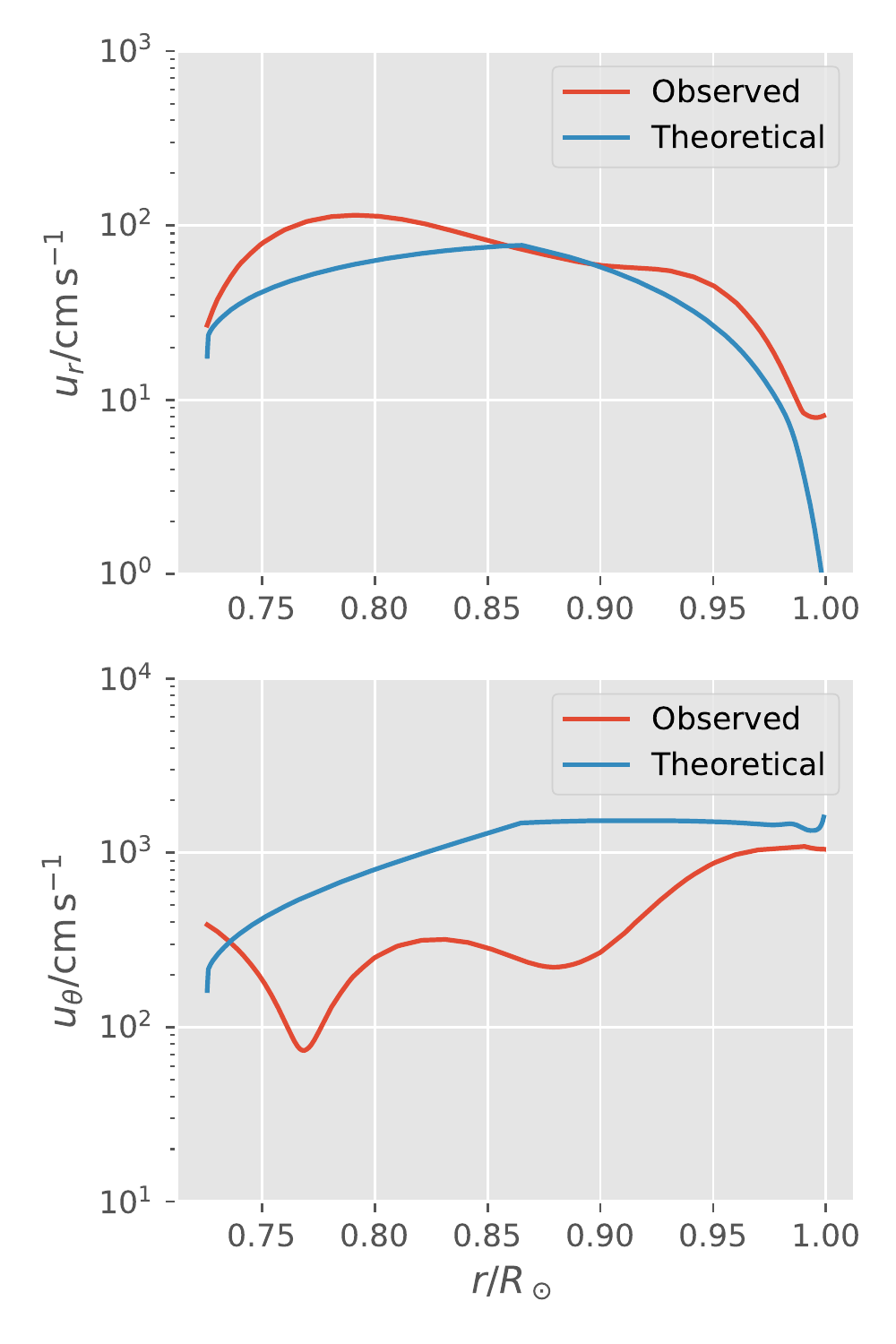}
\caption{The root mean square meridional velocity taken over latitudinal slices of the data is shown (red) for radial (top) and angular (bottom) components~\citep{2015ApJ...813..114R}. In blue are the predicted velocities with $\Omega\approx 2.5\times 10^{-6}\mathrm{Hz}$.}
\label{fig:vpre}
\end{figure}

As a test of this model, Fig.\ \ref{fig:vpre} shows the radial and latitudinal velocities,~\eqref{eq:ur} and~\eqref{eq:ut}, in the solar convection zone compared with the circulation velocity inferred from helioseismic observations~\citep{2015ApJ...813..114R}.
The solar model envelope, computed by G.~Houdek and D.~O.~Gough, is the model used by~\citet{2005MNRAS.360..859C} in their study of the power spectral density of solar p-modes.
We fitted the parameter $y$ so as to minimise the root mean square error in the logarithm of the velocity components.
This produced $y=0.2$, in agreement with the general magnitude we expect.
The variation in the radial component of the flow is well modelled by equation~\eqref{eq:ur} except near the surface, where the sharp density gradient means that inertial effects become increasingly important relative to thermal considerations.
It is therefore worth noting that equation~\eqref{eq:ur} is really a lower bound on the radial velocity set by the condition of thermal equilibrium: greater velocities are, of course, permitted.
The overall magnitude and trend in the $\theta$ component of the flow is reasonably captured by equation~\eqref{eq:ut}, but the details are not.
In particular, the dips in the observed $u_\theta$ are due to cell boundaries in the meridional flow structure, and these geometric features are not captured by our simplified analysis.
Furthermore we generally predict velocities which are larger than what is observed in deeper regions.
This could be a result of geometric or magnetic effects near the tachocline or else could indicate that we ought to have used a more precise prescription in equation~\eqref{eq:epsmax}.

A further test is provided by the simulations of~\citet{0004-637X-702-2-1078}, who find that in three-dimensional simulations of slowly rotating ($\Omega < |N|$) convecting giant stars of order $5$ to $10\,$per-cent of the kinetic energy resides in the meridional circulation, such that the circulation velocity is of order $30\,$per-cent of the convective velocity.
This is in good agreement with equations~\eqref{eq:ur} and~\eqref{eq:ut} when $\Omega > |N|$.

\section{Effects in Massive Stars}
\label{sec:massive}

In massive stars it is typical to have a convective core and a radiative envelope.
If the convection is anisotropic then the core is baroclinic and a meridional flow is present.
Here we argue that this produces baroclinicity in the radiative zone, which ultimately drives a circulation current there\footnote{This follows the standard Eddington--Sweet argument once we establish the magnitude of the baroclinicity.}.
We then examine the decay of this circulation into the radiative zone and examine how these effects scale in stellar models.

\subsection{Transmitting Baroclinicity}

To analyse the transmission of baroclinicity we begin by noting that in radiative zones
\begin{align}
	h s_r &= h \frac{\partial s}{\partial r}\\
	&= h \left[\frac{\partial \ln P}{\partial r} - \gamma \frac{\partial \ln \rho}{\partial r}\right]\\
	&= h \left[(1+\gamma)\frac{\partial \ln P}{\partial r} + \gamma \frac{\partial \ln T}{\partial r}\right]\\
	&= -(1+\gamma) + \gamma \frac{\mathrm{d}\ln T}{\mathrm{d}\ln P},
\end{align}
where we have taken the mean molecular weight $\mu$ to be constant and made use of equation~\eqref{eq:s} as well as the ideal gas law $P = \rho k_B T/\mu$.
In radiative zones $d \ln T / d \ln P$ is generally small compared to the adiabatic gradient, so $h s_r \approx -(1+\gamma)$ is of order unity.
Thus according to equation~\eqref{eq:lambda1}
\begin{equation}
	\lambda \approx \alpha.
	\label{eq:lalpha}
\end{equation}
At the tachocline, the boundary of the convection zone, we expect
\begin{equation}
	\alpha \approx P_\mathrm{tachocline}^y \int_{P_{\mathrm{tachocline}}}^{P_{\mathrm{c}}}\epsilon_{\mathrm{max}}\min\left(1, \frac{\Omega(\mathscr{P})^2}{|N|(\mathscr{P})^2}\right) \frac{\mathrm{d}\mathscr{P}}{\mathscr{P}^{1+y}},
\end{equation}
because $\alpha$ accumulates perturbations to the entropy gradient\footnote{It is important to note that it is $\alpha$, not $\alpha'$, which matters in radiative zones.
This is because the additional factor of $\Omega^2/|N|^2$ which arises through convective anisotropy vanishes in radiative zones.}.
In this way effects which would be small in the radiative zone, which is difficult to perturb because $|N|$ tends to be quite large there, are enhanced because they accumulate in the convection zone, where $|N|$ is small, and are then transmitted to the radiative zone at the tachocline.

To more precisely examine this transmission we write the heat flux as
\begin{align}
		\boldsymbol{F} = \boldsymbol{e}_r F_0(r) + A(r)\boldsymbol{e}_r \cos\theta + B(r) \boldsymbol{e}_{\theta} \sin\theta,
		\label{eq:fexp}
\end{align}
where
\begin{align}
	F_0 = \frac{L}{4\pi r^2}
\end{align}
is the unperturbed flux of the star, $L$ its luminosity and $A$ and $B$ are of order $\alpha F_0$ at the tachocline.
There is no $\cos\theta$ term along $\boldsymbol{e}_\theta$ because $\boldsymbol{e}_\theta \cdot \boldsymbol{F}$ must vanish for $\theta = 0$ and $\theta=\pi$, and we neglect higher-order harmonics for simplicity.
Within the radiative zone and in the absence of a meridional flow, the thermal flux obeys equation~\eqref{eq:diffflux}, which may be written as
\begin{equation}
	\boldsymbol{F} = -k \nabla T,
	\label{eq:old56}
\end{equation}
where the thermal conductivity
\begin{align}
	k \equiv \rho c_{p} \chi
\end{align}
depends solely on the temperature, pressure and opacity.
Hence
\begin{align}
		\nabla T = -\frac{1}{k}\left[\boldsymbol{e}_r F_0(r) + A(r)\boldsymbol{e}_r \cos\theta + B(r) \boldsymbol{e}_{\theta} \sin\theta\right].
		\label{eq:gradT}
\end{align}
It follows that when $\alpha$ is small
\begin{align}
	\frac{\partial_\theta T}{r \partial_r T} \approx \frac{B \sin \theta}{F_0},
\end{align}
which is of order $\alpha$.
Thus the heat flux transmits baroclinicity from the convection zone into the radiative zone to an extent given by equation~\eqref{eq:lalpha}.

Note that the {\em mechanical} transmission of currents across the tachocline has been studied in great detail by~\citet{2010ApJ...719..313G} and in the context of that work these stars are in a limit of efficient transmission, being neither mechanically nor thermally constrained\footnote{See Appendix~\ref{appen:diffrot} for further details.}.
Hence it is possible that mechanical pumping plays a significant role as well.

\subsection{Decay Profile}

Within the radiative zone the flux perturbation generally decays.
We now aim to determine the scale over which this occurs.
In equilibrium and in the absence of any meridional flow or heat generation, the flux obeys the conservation law
\begin{equation}
	\nabla\cdot\boldsymbol{F} = 0	
\end{equation}
and, by equation~\eqref{eq:old56},
\begin{equation}
	\nabla \times \boldsymbol{F} = -k\nabla\times\nabla T - \nabla k \times \nabla T = \nabla \ln k \times \boldsymbol{F}.
	\label{eq:curlF}
\end{equation}
From
\begin{align}
\nabla\cdot\boldsymbol{F}=0
\end{align}
and equation~\eqref{eq:fexp} we obtain
\begin{align}
	\label{eq:A}
	\frac{\partial A}{\partial r} + \frac{2}{r} (A+B) &= 0\\
	\intertext{and from equation\ \eqref{eq:curlF}}
	\label{eq:B}
	\frac{\partial B}{\partial r} + \frac{1}{r} (B-A) &= \frac{\partial \ln k}{\partial r} B - \frac{2}{\pi} \int_0^\pi \frac{\partial \ln k}{\partial \theta}\left[F_0 + \frac{1}{r}A(r) \cos\theta\right]\sin\theta d\theta\\
	&= \frac{\partial \ln k}{\partial r} B - \frac{2F_0}{\pi r} \int_0^\pi \frac{\partial \ln k}{\partial \theta}\sin\theta d\theta.
\end{align}
Because we have assumed a chemically homogeneous star we can relate $k$ in the limit of small $\alpha$ to the derivatives of $k$ in temperature and pressure.
That is,
\begin{align}
	\frac{\partial \ln k}{\partial \theta} \approx \frac{\partial \ln k}{\partial \ln T}\frac{\partial \ln T}{\partial \theta} +  \frac{\partial \ln k}{\partial \ln P}\frac{\partial \ln P}{\partial \theta},
	\label{eq:lnk1}
\end{align}
where the thermodynamic derivatives of $k$ are taken with respect to the unperturbed state and so are independent of $\theta$.
The pressure only acquires a dependence on $\theta$ through the centrifugal force, so
\begin{align}
	\frac{\partial \ln P}{\partial \theta} \approx \frac{\Omega^2 R}{g} \approx \frac{\Omega^2 r}{g} \sin\theta.
	\label{eq:lnk2}
\end{align}
The temperature depends on $\theta$ through equation~\eqref{eq:gradT} so
\begin{align}
	\frac{\partial \ln T}{\partial \theta} = -\frac{r B(r)}{k T}\sin\theta.
	\label{eq:lnk3}
\end{align}
Inserting equations~\eqref{eq:lnk1},~\eqref{eq:lnk2} and~\eqref{eq:lnk3} into equation~\eqref{eq:B} we find
\begin{align}
		\frac{\partial B}{\partial r} + \frac{1}{r} (B-A)&= \frac{\partial \ln k}{\partial r} B - \frac{F_0}{r} \left(-\frac{r B}{kT}\frac{\partial \ln k}{\partial \ln T} + \frac{\Omega^2 r}{g}\frac{\partial \ln k}{\partial \ln P}\right).
\end{align}
From equation~\eqref{eq:gradT} we find
\begin{align}
	\frac{F_0}{kT} \approx |\nabla \ln T| \approx -\frac{d\ln P}{dr} \frac{d\ln T}{d\ln P} = \frac{1}{h}\frac{d\ln T}{d\ln P}.
\end{align}
So
\begin{align}
		\frac{\partial B}{\partial r} + \frac{1}{r} (B-A)&= \left(\frac{\partial \ln k}{\partial r}-\frac{1}{h}\frac{d\ln T}{d\ln P}\frac{\partial \ln k}{\partial \ln T}\right) B - F_0 \frac{\Omega^2 r}{g}\frac{\partial \ln k}{\partial \ln P}.
\end{align}
The final term in this equation is important near the surface, where it is necessary to reproduce the usual Eddington--Sweet circulation.
However deeper in the star it may be neglected because there $B/F_0 \approx \alpha \gg \Omega^2 r/g$ and logarithmic derivatives of $k$ with respect to each of pressure and temperature are of order unity.
Hence we find
\begin{align}
		\frac{\partial B}{\partial r} + \frac{1}{r} (B-A)&= \left(\frac{\partial \ln k}{\partial r}-\frac{1}{h}\frac{d\ln T}{d\ln P}\frac{\partial \ln k}{\partial \ln T}\right) B,
\end{align}
which may be written as
\begin{equation}
	\frac{\partial B}{\partial r} + \frac{1}{r} (B-A) \approx -\frac{b}{h} B,
	\label{eq:B}
\end{equation}
where
\begin{align}
	b \equiv \frac{\mathrm{d}\ln k}{\mathrm{d}\ln P} + \frac{d\ln T}{d\ln P}\frac{\partial \ln k}{\partial \ln T}.
\end{align}

Combining equations~\eqref{eq:A} and~\eqref{eq:B} we find
\begin{align}
	\frac{\partial}{\partial r}\left(\begin{array}{c}
A\\
B
\end{array}\right)=\left(\begin{array}{cc}
-\frac{2}{r} & -\frac{2}{r}\\
+\frac{1}{r} & -\frac{1}{r}-\frac{b}{h}
\end{array}\right)\left(\begin{array}{c}
A\\
B
\end{array}\right).
\end{align}
Treating $b$ and $r$ as constants, the eigenvalues of this system are
\begin{align}
	\lambda_\pm = -\frac{1}{r}\left[\frac{3}{2} + \frac{br}{2h}\pm \frac{1}{2}\sqrt{\frac{b^2 r^2}{h^2} - 2b\frac{r}{h} - 7}\right].
	\label{eq:eigenvalues}
\end{align}
In and around the cores of stars $r$ is typically smaller than $h$, and $b$ is of order unity, so for simplicity we neglect terms involving $b$ and find that the slowest-decaying mode is
\begin{align}
	\lambda_- \approx -\frac{3}{2r} - i\frac{\sqrt{7}}{2r},
	\label{eq:eigen}
\end{align} 
so
\begin{align}
	\Re(\lambda) \approx -\frac{3}{2r}.
\end{align}
Taking only this mode, because it is the one that persists for the largest distance, we find that
\begin{align}
	\frac{\mathrm{d}\ln A}{\mathrm{d}\ln r} = \frac{\mathrm{d}\ln B}{\mathrm{d} \ln r} = \Re(\lambda). 
\end{align}
Hence
\begin{align}
	\frac{A(r)}{A(r_\mathrm{c})} \approx \frac{B(r)}{B(r_\mathrm{c})} \approx \left(\frac{P}{P_\mathrm{tachocline}}\right)^\beta,
\end{align}
where $r_\mathrm{c}$ is the radius of the convective core and
\begin{align}
	\beta \equiv -h \Re\left(\lambda_{-}\right) \approx \frac{3h}{2r}.
	\label{eq:beta}
\end{align}
Thus the baroclinicity is of order
\begin{equation}
	\alpha \approx \alpha(r_\mathrm{c}) \left(\frac{P}{P_\mathrm{tachocline}}\right)^\beta.
\end{equation}
This generally prevents solutions which do not involve $\mathcal{O}(\alpha)$ meridional flows for the same reason as outlined in section~\ref{sec:origins}.

It is worth noting that equation~\eqref{eq:eigen} indicates that the modes have an oscillatory component with characteristic scale $2r/\sqrt{7}$.
This suggests that the cell structure of the flow varies on that radial scale, though the level of detail in our calculations is not sufficient to determine more detail of how this manifests in the circulation pattern.

The standard Eddington--Sweet derivation for baroclinic radiative zones finds that the radial circulation velocity is of order that required to balance the flux anisotropy.
That is,
\begin{align}
	\rho c_p T u_r s_r \approx \nabla\cdot F
\end{align}
\citep{1929MNRAS..90...54E}.
Noting that $h s_r$ is of order unity we find
\begin{align}
	u_r \approx \frac{h \nabla\cdot F}{\rho c_p T}.
\end{align}
The length-scale associated with the flux anisotropy is $r$, because that is the latitudinal scale, and the flux scale of the anisotropy is set by $\alpha F$, so
\begin{align}
	u_r \approx \alpha \frac{h F}{r \rho c_p T}.
\end{align}
Using equation~\eqref{eq:urt} we find
\begin{equation}
u \approx \alpha \frac{F}{\rho c_p T}.
\label{eq:uradiative}
\end{equation}
In an efficient convection zone the luminosity equals the convective luminosity.
This is well approximated by the power flux of a fluid moving with the convection speed, so
\begin{align}
	L \approx 4\pi r^2 \rho u_c^3,
	\label{eq:lumvc}
\end{align}
where $u_\mathrm{c}$ is the convection speed, the heat flux outside of the core may be related to that at the boundary of the core by
\begin{equation}
	F \approx \rho_{\mathrm{c}} u_{\mathrm{c,core}}^3 \left(\frac{r_\mathrm{c}}{r}\right)^2.
\end{equation}
and so
\begin{align}
	\frac{u_r}{u_\mathrm{c,core}} \approx\  &\left(\frac{\rho_\mathrm{c}}{\rho}\right)\left(\frac{u_\mathrm{c,core}}{c_\mathrm{s}}\right)^2\left(\frac{r_\mathrm{c}}{r}\right)^2\left(\frac{P}{P_\mathrm{tachocline}}\right)^\beta \alpha(r_\mathrm{c}),
	\label{eq:vmass}
\end{align}
where $c_\mathrm{s}$ is the sound speed and
\begin{equation}
	\alpha(r_\mathrm{c}) = P_{\mathrm{tachocline}}^{y}\int_{P_{\mathrm{tachocline}}}^{P_{\mathrm{c}}}\epsilon_{\mathrm{max}}\min\left(1, \frac{\Omega(\mathscr{P})^2}{|N|(\mathscr{P})^2}\right) \frac{\mathrm{d}\mathscr{P}}{\mathscr{P}^{1+y}}.
\end{equation}

It is worth noting that the ratio $P/P_\mathrm{tachocline}$ scales exponentially in the radial coordinate while the flow speed scales as a power law in this ratio.
The latter scaling persists even after accounting for the additional heat flux transported by the meridional flow.
To see this note that the rate at which $\alpha$ changes with $r$ is proportional to the flux transported latitudinally by the meridional flow because this flow damps $\alpha$.
This flux scales as $\alpha$ because the flow transports material radially at a rate which scales as $\alpha$ and the radial gradient is largely independent of $\alpha$.
This means that $d\alpha/dr \approx -(h/r^2)\alpha$, which produces a power-law in $P$, which merely modifies the exponent $\beta$.
The modification is approximately
\begin{align}
	\beta \rightarrow \beta + \frac{h}{r},
\end{align}
applied to equation~\eqref{eq:beta}, so that
\begin{align}
	\beta \approx \frac{5h}{2r}.
	\label{eq:beta1}
\end{align}

As one final simplification, it is worth noting that when the exponent $\beta$ is unity equation~\eqref{eq:uradiative} may be written as
\begin{equation}
	u_r \approx  \left(\frac{F}{P_\mathrm{c}}\right)\alpha_\mathrm{core}.
\end{equation}
This form has the disadvantage of not generalising should the exponent differ from one and of lacking the clear dimensionless ratios of equation~\eqref{eq:vmass}, but has the advantage that the origin of the flow is clear: rotationally-driven ansiotropy forces a meridional flow to carry a portion of the flux, with the speed set by the energy density $P$.

\subsection{Scaling in Stellar Models}

\begin{figure}
	\includegraphics[clip,width=\columnwidth]{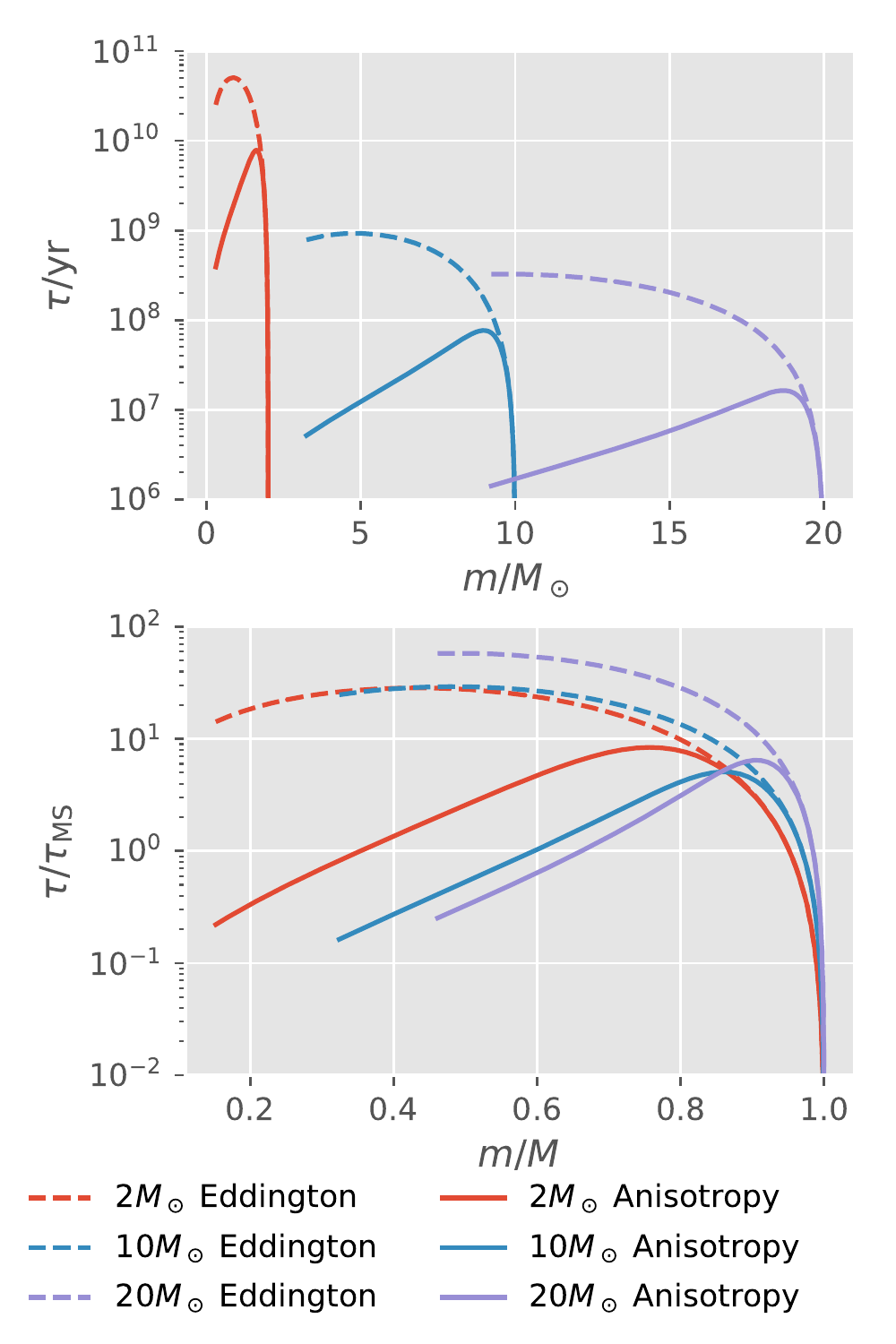}%
	\caption{
The analogue of the Eddington--Sweet time $\tau = z / u_r$ is shown as a function of mass $m$ for three stellar models on the main sequence with masses of $2M_\odot$, $10M_\odot$ and $20M_\odot$.
For each model results from  both the Eddington--Sweet mechanism and the mechanism presented in this paper are shown.
These calculations were made with $\beta=3/2$ and rigid rotation with a surface velocity of $u_\mathrm{rot} = 3\times 10^6\,\mathrm{cm\,s^{-1}}$.
The bottom panel shows the same timescales $\tau$ normalised to the main-sequence lifetime $\tau_\mathrm{MS}$, with the mass normalised to the total mass of the star $M$.}
	\label{fig:times}
\end{figure}

To understand these effects quantitatively we define
\begin{align}
	\tau \equiv \frac{h}{u_r}
\end{align}
as the analogue of the local Eddington--Sweet time.
Noting that $h$ is approximately the distance $z = R - r$ to the surface of the star we see that $\tau$ is an estimate of the time needed to bring material to the surface.
Fig.~\ref{fig:times} shows this time as a function of mass coordinate for three different stellar models made with the Cambridge STARS code~\citep{1971MNRAS.151..351E,1995MNRAS.274..964P}.
These calculations were made with a surface rotation of $u_\mathrm{rot} = 3\times 10^6\,\mathrm{cm\,s^{-1}}$ and we assume that the mixing rate cannot fall below that of the Eddington circulation.
For simplicity we took $\beta=3/2$ though in practice this parameter varies with $r$.

A few implications are clear from this figure.
First, rotating massive stars mix near the core much more rapidly than would normally be expected without accounting for convective anisotropy, which increases the effective core mass and acts very much like convective overshooting~\citep{1974ApJ...193..109P}, a topic we explore in more detail in Section~\ref{sec:overshoot}.
This enhancement arises because $|N|$ in the core is generally much smaller than $\sqrt{g/R}$, and so the convective boundary condition drives baroclinicity much more than the simple Eddington--Sweet mechanism.
Secondly, this effect damps strongly as we approach the surface and is eventually overtaken by the Eddington circulation, so there would not necessarily be strong observable chemical signatures of the enhanced mixing except by virtue of making material more readily available to other mixing processes.

\begin{figure}
	\includegraphics[clip,width=\columnwidth]{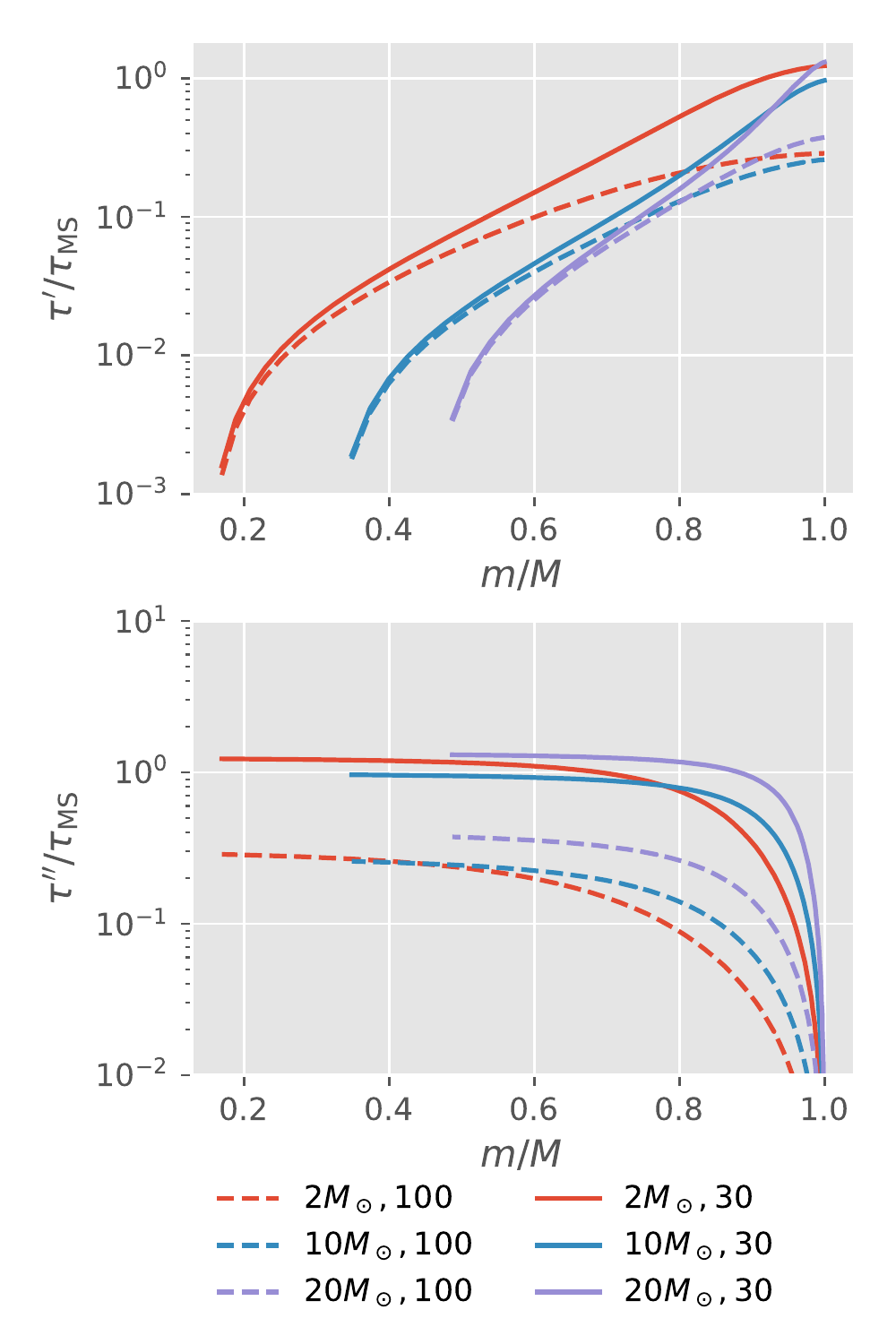}%
	\caption{
The time $\tau'$ required for material to flow to the core (top) and the time $\tau''$ required for material to flow to the surface (bottom) are shown as functions of mass fraction $m/M$ for three stellar models on the main sequence with masses of $2M_\odot$, $10M_\odot$ and $20M_\odot$.
Both times are normalised to the main-sequence lifetime $\tau_{\rm MS}$.
For each model the velocity $u_r$ is assumed to be the sum of that owing to the Eddington--Sweet mechanism and that owing to the mechanism presented in this paper.
These calculations were made with $\beta=3/2$ and rigid rotation with a surface velocities of $u_\mathrm{rot} = 3\times 10^6\,\mathrm{cm\,s^{-1}}$ (solid) and $u_\mathrm{rot} = 10^7\,\mathrm{cm\,s^{-1}}$ (dashed). The former correspond respectively to $4\times 10^{-3}$, $2\times 10^{-3}$ and $10^{-3}$ of the surface breakup rate for the three stars, while the latter correspond to $4\times 10^{-2}$, $2\times 10^{-2}$ and $10^{-2}$ of the breakup rate respectively.}
	\label{fig:times2}
\end{figure}

To understand how effectively this mechanism transports material through the star we now examine the travel time from the tachocline to a point at radius $r$, given by
\begin{align}
	\tau' \equiv \int_{r_\mathrm{c}}^r \frac{dr}{u_r}.
\end{align}
This is shown in Figure~\ref{fig:times2}.
In each of the stellar models in Fig.~\ref{fig:times} $\tau'$ is less than the lifetime of the star almost everywhere and so, at least at this rotation rate, a significant amount of material ought to reach the core.
By contrast, the time
\begin{align}
	\tau'' \equiv -\int_{r_\mathrm{surface}}^r \frac{dr}{u_r}.
\end{align}
to mix to the surface exceeds the main-sequence lifetime in most of the star.
Hence at least at this rotation rate the amount of material mixed from the core to the surface from this region ought to be minimal.
For more rapidly rotating stars this effect is stronger, principally because the Eddington--Sweet circulation is stronger, and considerably more material may reach the surface from the core.

\begin{figure}
	\includegraphics[clip,width=\columnwidth]{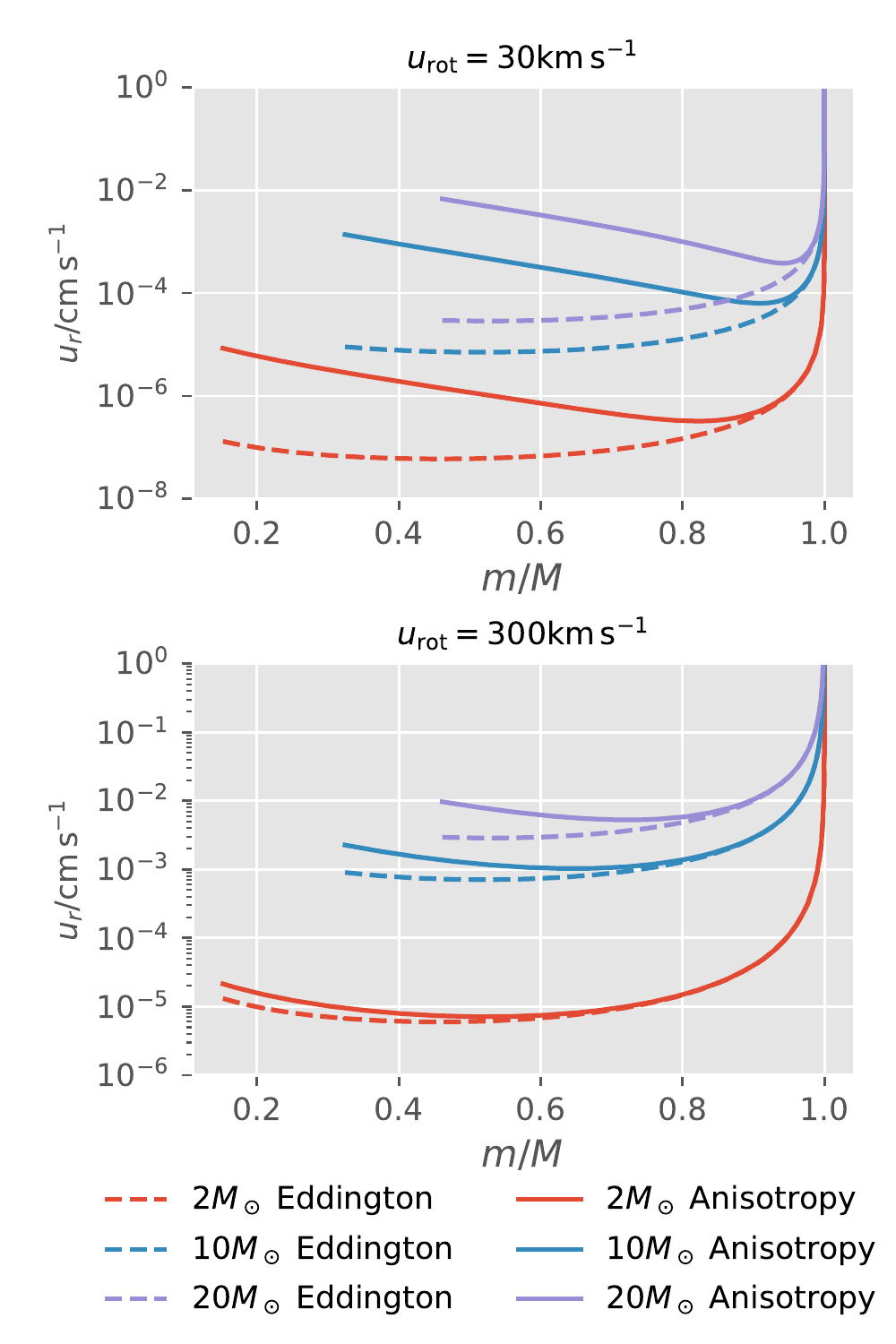}%
	\caption{
The radial velocity $u_r$ is shown as a function of mass fraction $m/M$ for three stellar models on the main sequence with masses of $2M_\odot$, $10M_\odot$ and $20M_\odot$.
For each model results from  both the Eddington--Sweet mechanism and the mechanism presented in this paper are shown.
These calculations were made with $\beta=3/2$ and for simplicity assume rigid rotation.
The top panel has surface velocity of $u_\mathrm{rot} = 3\times 10^6\,\mathrm{cm\,s^{-1}}$, corresponding respectively to $4\times 10^{-3}$, $2\times 10^{-3}$ and $10^{-3}$ of the surface breakup velocity for the three stars. The bottom panel has $u_\mathrm{rot} = 3\times 10^7\,\mathrm{cm\,s^{-1}}$, corresponding respectively to $0.4$, $0.2$ and $0.1$ of the breakup velocity.}
	\label{fig:D}
\end{figure}

A similar story may be seen in Fig.~\ref{fig:D}, which shows the radial velocity of the circulation as a function of position in the star.
The results are for the same stellar models as in Fig.~\ref{fig:times} and, once more, the Eddington curves only contain the Eddington--Sweet circulation while the Anisotropy curves contain both effects.
Because the anisotropy-induced circulation declines towards the surface it is clear from this where the Eddington--Sweet mechanism becomes dominant.
In the top panel, which was made with a surface velocity of $u_\mathrm{rot} = 30\,\mathrm{km\,s^{-1}}\,$, this point is typically around a mass fraction of $0.9$.
By contrast in the bottom panel, which was made with a surface velocity of $u_\mathrm{rot} = 300\,\mathrm{km\,s^{-1}}\,$, the transition occurs between $0.5$ and $0.7$.
This is because the anisotropy-driven circulation saturates at much lower rotation rates and so does not scale with rotation in this regime while the Eddington--Sweet mechanism scales quadratically with rotation rate.

It is worth noting that a similar effect is possible in stars of lower mass, where the anisotropy in the outer convection zone induces baroclinicity in the radiative core.
This effect is more strongly damped because the characteristic damping scale is $h$, which is much shorter at an outer radiative--convective boundary than at a core convective--radiative boundary.
Equivalently, the damping goes as $P_\mathrm{tachocline}/P$ in this case, and $P_\mathrm{tachocline}$ may be quite small relative to the core pressure.
In addition the velocity damps towards the core because the temperature increases, making it easier to dissipate the flux accumulation.
Finally, the anisotropy for a given rotation is smaller because the relevant $|N|$ is greater.
Still, in the Sun $h \approx R_{\sun}/10$ near the tachocline, so this likely causes some mixing between the convection zone and the material a few tenths of a solar radius below it.

\section{Chemical Composition Gradients}
\label{sec:chemistry}

In the convection zones of massive stars chemical composition gradients are generally wiped out by turbulent mixing.
So there are significant composition gradients only in the radiative zones of these bodies.

Because the circulation in radiative zones is driven by the same criterion of thermal equilibrium as the Eddington--Sweet circulation such gradients act on it in the same manner as they do that mechanism.
The only place where there might be a difference in how composition gradients impact the mechanism we have introduced is in the transmission of baroclinicity from the convection zone to the radiative zone.
However this transmission is achieved by the core emitting an aspherical heat flux and this is not at all changed by the chemistry in the vicinity of the core.
Hence the effect of composition ought to be the same in both mechanisms.
This has been examined in detail by~\citet{2000A&A...361..101M}, who provide a prescription for correcting the circulation velocity and find that such corrections are typically of order unity.
This is therefore what we expect.

\section{Convective Overshooting}
\label{sec:overshoot}

The fact that these circulations drive extra mixing in radiative zones close to convective cores suggests that this may be the source of extra mixing which is commonly, though probably erroneously, called convective overshooting or penetration~\citep{1974ApJ...193..109P}.
Such extra mixing naturally prolongs the main-sequence lifetime as required to match a number of evolved binary systems~\citep{1997MNRAS.285..696S, 2015ApJ...807..184F,2016arXiv161105041E} and so is of significant interest for the purposes of stellar modelling.

Convective overshooting is often incorporated in stellar models by modelling a region beyond the convection zone as isentropic and well-mixed.
This region has width
\begin{align}
	l_{\mathrm{ov}} = \alpha_{\mathrm{ov}} h,
	\label{eq:lov}
\end{align}
where $\alpha_{\mathrm{ov}}$ is a dimensionless parameter\footnote{Confusingly, this is also called $f_{\mathrm{ov}}$ by some authors.}~\citep{1997MNRAS.285..696S}.
A long-standing problem with this prescription is that the overshoot distance $\alpha_{\mathrm{ov}} h$ required to match observations is often much larger than what calculations of the stiffness of the radiative-convective boundary suggest~\citep{1965ApJ...142.1468S}.
What appears more likely is that the physical overshooting is small as predicted by stiffness calculations but that there is additional mixing caused by the baroclinic mechanism developed in the previous section.

The most immediate comparison with observations that can be made to test this hypothesis is to calculate $\alpha_{\mathrm{ov}}$ from our model.
To do so we calculate $l_{\mathrm{ov}}$ such that in the lifetime of the star the material within this distance of the tachocline is well-mixed with that at the tachocline.
This is just the statement that
\begin{align}
	\frac{l_{\mathrm{ov}}}{u_r} \approx \tau_\mathrm{MS},
\end{align}
or
\begin{align}
	l_{\mathrm{ov}} \approx u_r \tau_\mathrm{MS},
	\label{eq:orig}
\end{align}
where $\tau_\mathrm{MS}$ is the main-sequence lifetime of the star.
This is not quite right, however, because vertical chemical mixing is less efficient than the circulation velocity alone suggests.
As suggested by~\citet{1998A&A...334.1000M} the effective mixing is reduced by a factor of
\begin{align}
	w \equiv \frac{r u_r}{30 D_{\rm h}},	
\end{align}
where $D_{\rm h}$ is the horizontal diffusivity.
From Appendix~\ref{appen:diffrot} we have
\begin{align}
	D_{\rm h} \approx \nu \approx \frac{T}{\rho S},
\end{align}
where $\nu$ is the turbulent viscosity, $S$ is the shear owing to differential rotation and $T$ is the magnitude of the turbulent stress.
With equation~\eqref{eq:nou} we find
\begin{align}
	D_{\rm h} \approx \frac{u h \Omega}{S}.
\end{align}
Hence
\begin{align}
	w \approx \frac{r u_r}{30 u h}\left(\frac{S}{\Omega}\right).	
\end{align}
From equation~\eqref{eq:urt} we find
\begin{align}
	w \approx \frac{1}{30}\left(\frac{S}{\Omega}\right).	
\end{align}
In Appendix~\ref{appen:diffrot} we found $S \ga \Omega$.
For simplicity, and as a reflection of the uncertainties involved in determining $S$, we use
\begin{align}
	w \approx \frac{1}{30}.
\end{align}
Hence equation~\eqref{eq:orig} becomes
\begin{align}
	l_{\mathrm{ov}} \approx \frac{1}{30}u \tau_\mathrm{MS}.
	\label{eq:orig2}
\end{align}

\begin{figure}
	  \includegraphics[clip,width=\columnwidth]{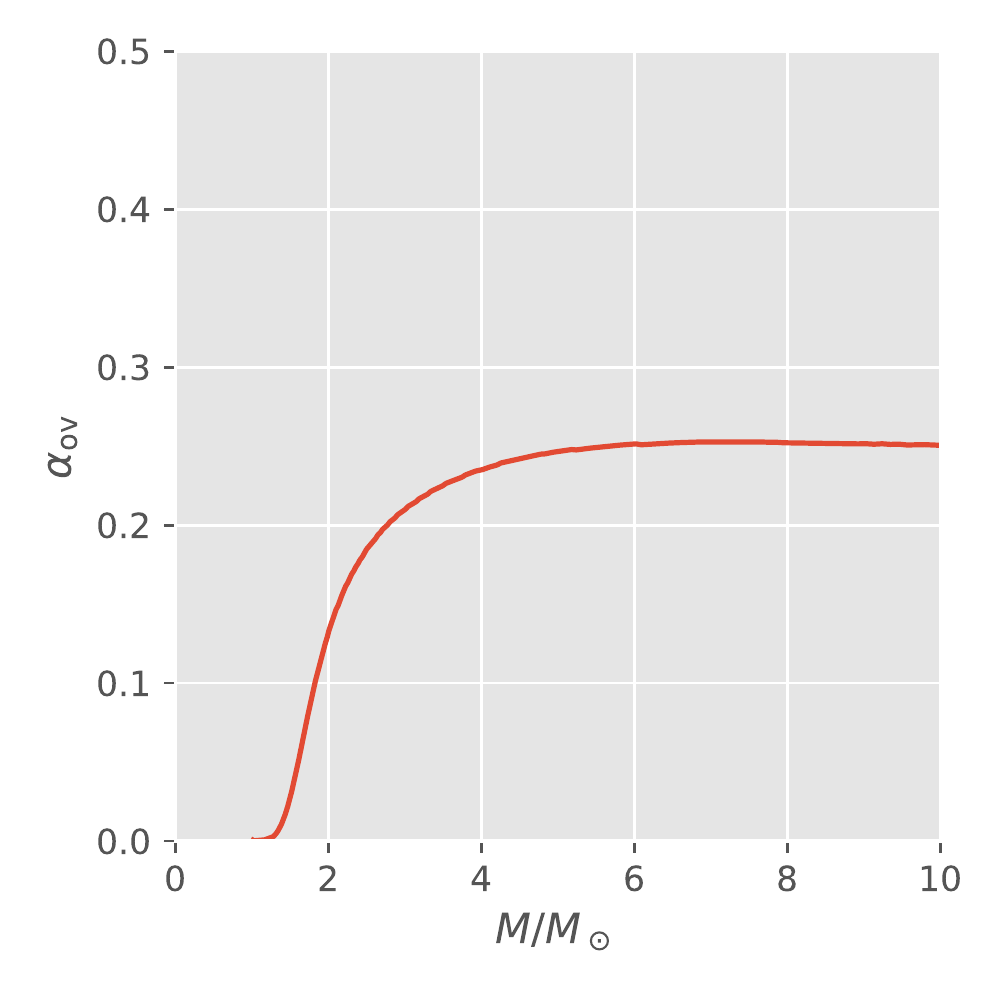}%
	\caption{The convective overshoot parameter $\alpha_{\mathrm{ov}}$ is shown as a function of stellar mass, computed with equation~\eqref{eq:delta} and $\epsilon_{\mathrm{max}} \approx 0.2$ as various turbulence closure schemes suggest~\citep{2013IAUS..294..399K,jermyn}.}
	\label{fig:delta}
\end{figure}

Inserting equation~\eqref{eq:vmass} into equation~\eqref{eq:orig2} we find
\begin{align}
	l_{\mathrm{ov}} &\approx \frac{u_{\mathrm{c,core}}}{30} \tau_\mathrm{MS}\left(\frac{\rho_\mathrm{c}}{\rho}\right)\left(\frac{u_\mathrm{c,core}}{c_\mathrm{s}}\right)^2\left(\frac{r_\mathrm{c}}{r}\right)^2\left(\frac{P}{P_\mathrm{tachocline}}\right)^\beta\alpha_\mathrm{core}.
\end{align}
In massive stars with even mild rotation, $\Omega > |N|$ everywhere in the core so
\begin{align}
	\alpha_\mathrm{core} &= P_{\mathrm{tachocline}}\int_{P_{\mathrm{tachocline}}}^{P_{\mathrm{c}}}\min\left(\epsilon_{\mathrm{max}},\frac{\Omega(\mathscr{P})^2}{|N|(\mathscr{P})^2}\right) \frac{d\mathscr{P}}{\mathscr{P}^2}\\
	&=\epsilon_{\mathrm{max}}P_{\mathrm{tachocline}} \int_{P_{\mathrm{tachocline}}}^{P_{\mathrm{c}}}\frac{d\mathscr{P}}{\mathscr{P}^2}\\
	&\approx \epsilon_{\mathrm{max}},
\end{align}
because $P_{\mathrm{tachocline}} \ll P_\mathrm{c}$.
Hence
\begin{align}
	l_{\mathrm{ov}} &\approx \frac{\epsilon_{\mathrm{max}}}{30}u_{\mathrm{c,core}} \tau_\mathrm{MS}\left(\frac{\rho_\mathrm{c}}{\rho}\right)\left(\frac{u_\mathrm{c,core}}{c_\mathrm{s}}\right)^2\left(\frac{r_\mathrm{c}}{r}\right)^2 \left(\frac{P}{P_\mathrm{tachocline}}\right)^\beta\\
	&\approx \frac{\epsilon_{\mathrm{max}}}{30}u_{\mathrm{c,core}} \tau_\mathrm{MS}\left(\frac{u_\mathrm{c,core}}{c_\mathrm{s, core}}\right)^2\left(\frac{r_\mathrm{c}}{r}\right)^2 \left(\frac{P}{P_\mathrm{tachocline}}\right)^{\beta-1},
\end{align}
where we have used the fact that $P \propto \rho c_{\mathrm{s}}^2$.
Noting that near the tachocline
\begin{align}
	r \approx r_\mathrm{c}
\end{align}
and
\begin{align}
	P \approx P_\mathrm{tachocline} e^{-\delta r/h},
\end{align}
where $\delta r$ is the distance to the tachocline, we find that
\begin{align}
	l_{\mathrm{ov}} &\approx \frac{\epsilon_{\mathrm{max}}}{30}u_{\mathrm{c,core}} \tau_\mathrm{MS}\left(\frac{u_\mathrm{c,core}}{c_\mathrm{s,core}}\right)^2 e^{-l_{\mathrm{ov}}(\beta-1)/h}.
\end{align}
Inserting equation~\eqref{eq:lov} we find
\begin{align}
	\alpha_{\mathrm{ov}} &\approx \frac{\epsilon_{\mathrm{max}}}{30}\frac{u_{\mathrm{c,core}}}{h}  \tau_\mathrm{MS}\left(\frac{u_\mathrm{c,core}}{c_\mathrm{s,core}}\right)^2 e^{-\alpha_{\mathrm{ov}}(\beta-1)}
\end{align}
and making use of equation~\eqref{eq:beta1}
\begin{align}
	\alpha_{\mathrm{ov}} &\approx \frac{\epsilon_{\mathrm{max}}}{30}\frac{u_{\mathrm{c,core}}}{h}  \tau_\mathrm{MS}\left(\frac{u_\mathrm{c,core}}{c_\mathrm{s,core}}\right)^2 e^{-\alpha_{\mathrm{ov}}(5h/2r -1)}.
	\label{eq:delta}
\end{align}
This may be solved in terms of the Lambert~W function or else numerically.
Fig.~\ref{fig:delta} shows $\alpha_{\mathrm{ov}}$ as a function of stellar mass for a fine grid of stellar models made with the Cambridge STARS code~\citep{1971MNRAS.151..351E,1995MNRAS.274..964P}.
We have taken $\epsilon_{\mathrm{max}}$ to be $0.2$ as suggested by various turbulence closure schemes~\citep{2013IAUS..294..399K,jermyn}.
The stellar lifetime was computed with the main-sequence fit of~\citet{1989ApJ...347..998E}.
The variation of $\alpha_{\mathrm{ov}}$ with mass suggests that it rises rapidly in the window between $1$ and $2\,\mathrm{M}_\odot$ and then asymptotically, shortly thereafter, $\alpha_{\mathrm{ov}} \approx 0.25$.
The variation is primarily driven by variation in $h/r_{\rm c}$ because the convective Mach number varies little with mass.
This behaviour is in very good agreement with the findings of~\citet{2000MNRAS.318L..55R},~\citet{2016ApJ...823..130M} and~\citet{2015A&A...575A.117S}, while yielding a slight overestimate relative to those of~\citet{0004-637X-849-1-18}, who find an asymptote of $\alpha_{\mathrm{ov}} \approx 0.2$.

Note that even very slowly rotating stars have $\Omega > |N|$, as $|N| \la 10^{-8}{\rm s^{-1}}$ in core convection zones.
For instance KIC~10526294, which rotates with a period of $188{\rm d}$ and hence has $\Omega \approx 4\times 10^{-7}{\rm s^{-1}}$, is a sufficiently rapid rotator that we expect it to follow equation~\eqref{eq:delta} and indeed this is approximately what is seen~\citep{2015A&A...580A..27M}.
For stars with rotation periods in excess of thousands of days it might be possible to observe the quadratic dependence of circulation $\Omega$ that we predict, though such objects are quite rare.

An alternate way to parametrise convective overshooting is as an additional diffusivity in the radiative zone of the form
\begin{align}
	D_\mathrm{overshoot} = D_{\mathrm{convective}} e^{-2\delta r/(f_{\mathrm{ov}} h)}
	\label{eq:overshoot}
\end{align}
\citep{1997A&A...324L..81H,2011ApJS..192....3P}, where $f_{\mathrm{ov}}$ is the exponential overshoot parameter, $\delta r$ is the distance to the radiative-convective boundary and
\begin{align}
	D_{\mathrm{convective}} \equiv u_{\mathrm{c, core}} h.
	\label{eq:dconv}
\end{align}
There are two ways to compare this prescription with our calculations.
First, we may note from a practical perspective that the $f_{\mathrm{ov}}$ and $\alpha_{\mathrm{ov}}$ prescriptions agree with one another, and so the fact that our calculated $\alpha_{\mathrm{ov}}$ are consistent with those inferred from observations implies consistency with the $f_{\mathrm{ov}}$ prescription.
Secondly, we can calculate an effective $f_{\mathrm{ov}}$ from our model.
Because we predict different spatial behaviour from that of equation~\eqref{eq:overshoot} there is no unique way to do this and different prescriptions yield different results.
Indeed to self-consistently map from mixing via circulation currents to an exponential convective overshoot model requires a detailed comparison of stellar models with each prescription which we leave for future work.
Nevertheless one straightforward way to perform this mapping is to insist that one scale-height from the convective-radiative boundary, the effective diffusivity owing to the meridional circulation
\begin{align}
	D_\mathrm{merid} &\approx u h
	\label{eq:merid}
\end{align}
equals the overshoot diffusivity.
Following the same procedure as before to calculate $u$ and incorporating the correction to chemical mixing we find
\begin{align}
	D_\mathrm{merid}&\approx \frac{\epsilon_{\mathrm{max}}}{30}u_{\mathrm{c,core}} h \left(\frac{u_\mathrm{c,core}}{c_\mathrm{s,core}}\right)^2\left(\frac{r_\mathrm{c}}{r}\right)^2 \left(\frac{P}{P_\mathrm{tachocline}}\right)^{\beta-1}.
\end{align}
Letting $D_\mathrm{merid}=D_\mathrm{overshoot}$ and inserting equation~\eqref{eq:dconv} we obtain
\begin{align}
	e^{-2 \delta r/(f_{\mathrm{ov}} h)}&\approx \frac{\epsilon_{\mathrm{max}}}{30}\left(\frac{u_\mathrm{c,core}}{c_\mathrm{s,core}}\right)^2\left(\frac{r_\mathrm{c}}{r}\right)^2 \left(\frac{P}{P_\mathrm{tachocline}}\right)^{\beta-1}.
\end{align}
Noting that near the tachocline $r \approx r_{\mathrm{c}}$ and
\begin{align}
	P \approx P_\mathrm{tachocline} e^{-\delta r/h},
\end{align}
we find
\begin{align}
	e^{-2 \delta r/(f_{\mathrm{ov}} h)}&\approx \frac{\epsilon_{\mathrm{max}}}{30} e^{-(\beta-1) \delta r/h}\left(\frac{u_\mathrm{c,core}}{c_\mathrm{s,core}}\right)^2.
\end{align}
Hence
\begin{align}
	\left[\frac{2}{f_{\mathrm{ov}}} - (\beta-1)\right]\frac{\delta r}{h} = \ln\frac{30}{\epsilon_{\mathrm{max}}}-2\ln\frac{u_\mathrm{c,core}}{c_\mathrm{s,core}}.
\end{align}
Letting $\delta r = h$ we find
\begin{align}
	\frac{2}{f_{\mathrm{ov}}} - (\beta-1) = \ln\frac{30}{\epsilon_{\mathrm{max}}}-2\ln\frac{u_\mathrm{c,core}}{c_\mathrm{s,core}}.
	\label{eq:alpha00}
\end{align}
Once more inserting equation~\eqref{eq:beta1} and taking $r \approx r_{\mathrm{c}}$ we find
\begin{align}
	\frac{2}{f_{\mathrm{ov}}} - \left(\frac{5h}{2r_\mathrm{c}} - 1\right) = \ln\frac{30}{\epsilon_{\mathrm{max}} }-2\ln\frac{u_\mathrm{c,core}}{c_\mathrm{s,core}}.
	\label{eq:alpha1}
\end{align}
Finally solving for $f_{\rm ov}$ yields
\begin{align}
	f_{\rm ov} = \frac{2}{\ln\frac{30}{\epsilon_{\mathrm{max}} }-2\ln\frac{u_\mathrm{c,core}}{c_\mathrm{s,core}} + \left(\frac{5h}{2r_\mathrm{c}} - 1\right)}.
\end{align}

\begin{figure}
	  \includegraphics[clip,width=\columnwidth]{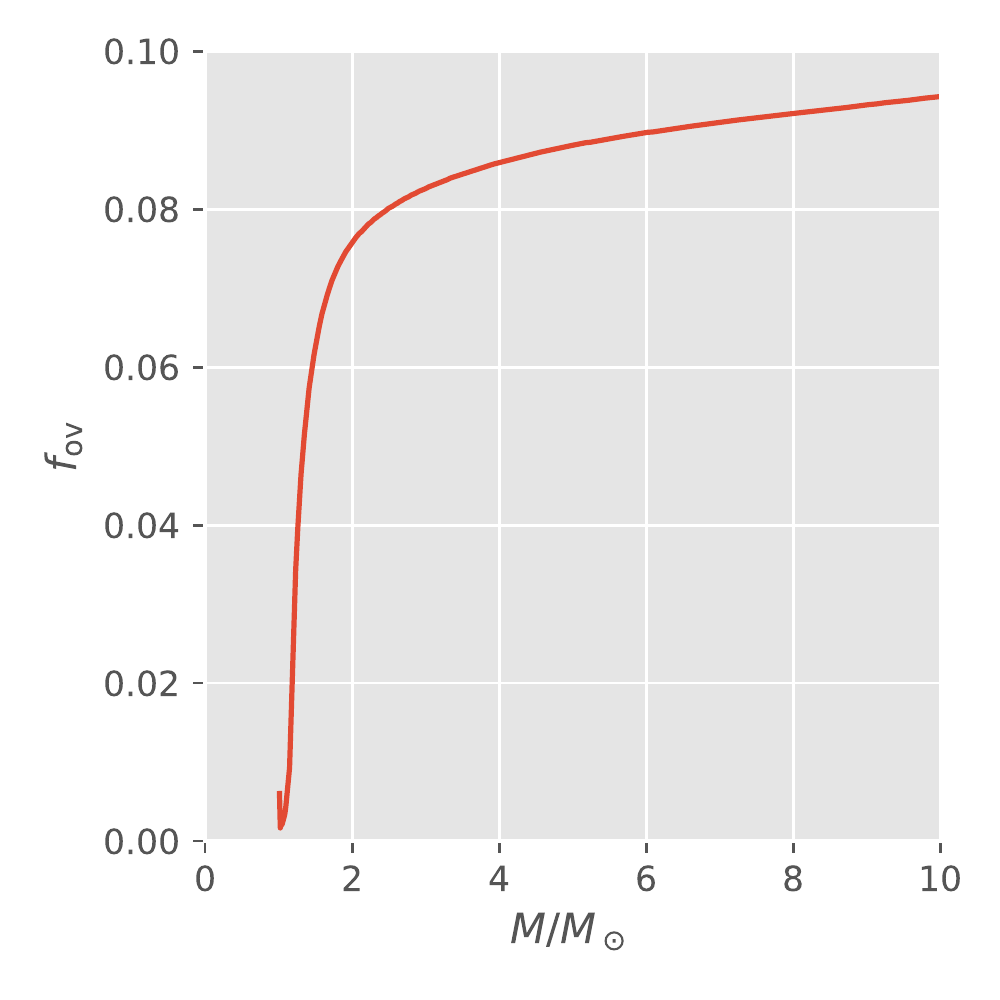}
	\caption{The convective overshoot parameter $f_{\mathrm{ov}}$ is shown as a function of stellar mass, computed with equation~\eqref{eq:alpha1}. The input stellar models were the same as those in Figure~\ref{fig:delta}.}
	\label{fig:alpha}
\end{figure}

Fig.~\ref{fig:alpha} shows $f_{\mathrm{ov}}$ as a function of stellar mass with the same stellar models and parameters as were used for Fig.~\ref{fig:delta}.
The results qualitatively capture the inferred dependence on stellar mass, but we overestimate $f_{\mathrm{ov}}$ by a factor of approximately $2$~\citep{2015A&A...575A.117S,2016ApJ...823..130M}.
This is not surprising given the ad-hoc nature of our matching between equations~\eqref{eq:overshoot} and~\eqref{eq:merid} but it is encouraging that the overall magnitude is approximately correct and that the dependence on mass reproduces what is observed.

\begin{figure}
	\includegraphics[clip,width=\columnwidth]{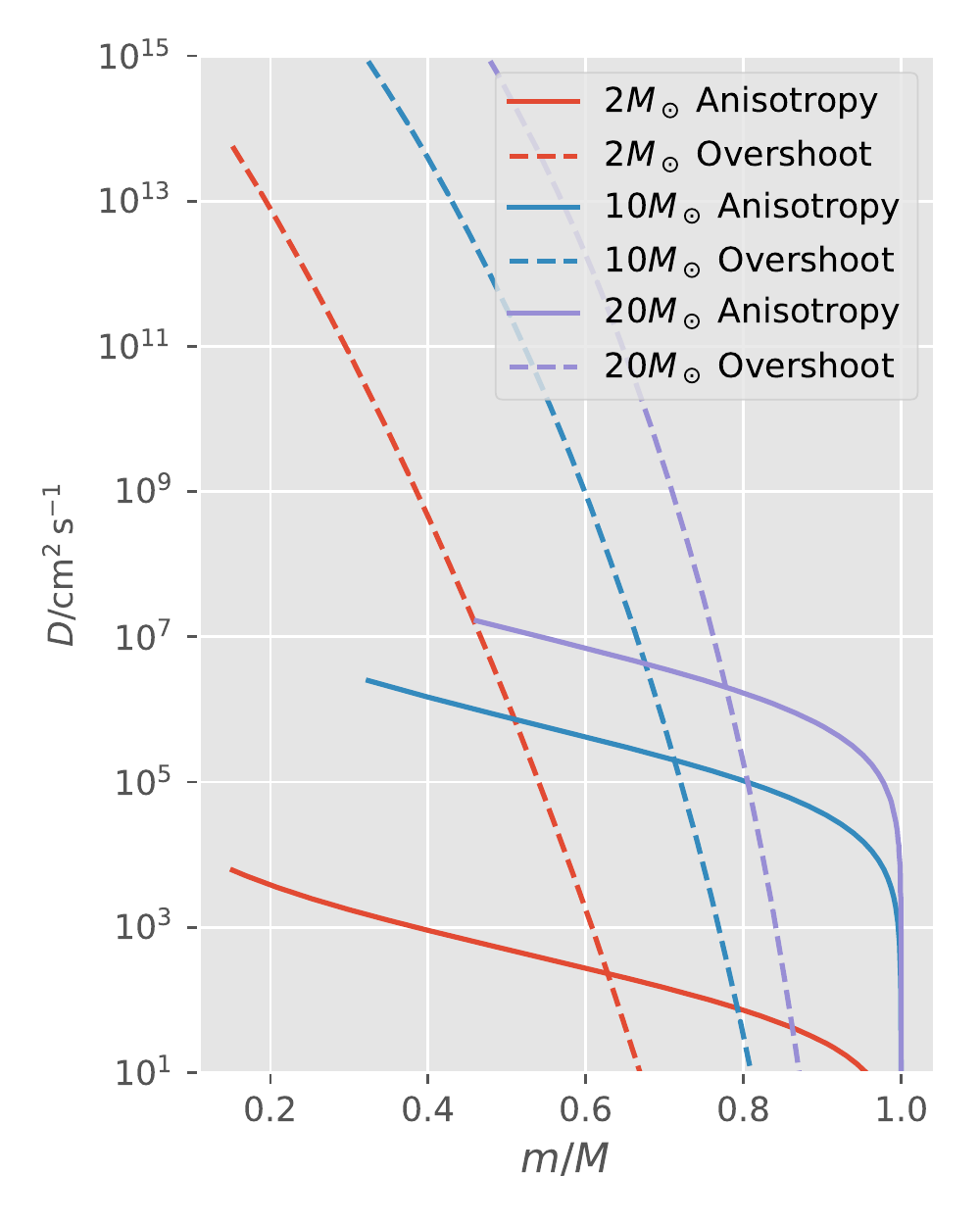}%
	\caption{
The effective circulation chemical diffusivity $D = h u_r / 30$ and convective overshoot diffusivity are shown as functions of mass fraction $m/M$ for three stellar models on the main sequence with masses of $2M_\odot$, $10M_\odot$ and $20M_\odot$.
The circulation diffusivity was calculated with $\beta = 3/2$ and only includes that owing to convective anisotropy and not the Eddington--Sweet mechanism.
The overshoot diffusivity was modelled with an exponential with $f_\mathrm{ov}=0.09$.
These calculations assume rigid rotation and a surface velocity of $u_\mathrm{rot} = 30\,\mathrm{km\,s^{-1}}$.}
	\label{fig:D2}
\end{figure}

As a check of these calculations, Fig.~\ref{fig:D2} shows the spatial dependence of both the circulation diffusivity and the convective overshoot diffusivity for the three stellar models considered for Fig.~\ref{fig:times}.
The overshoot diffusivity was calculated with the exponential parametrisation and $f_\mathrm{ov} = 0.09$, which is representative from Fig.~\ref{fig:alpha}.
Note that, in each case, the crossover occurs roughly one scale-height from the edge of the convective core, in keeping with our matching to the exponential parametrisation.
This both confirms the results of equation~\eqref{eq:alpha1} and Fig.~\ref{fig:alpha} and demonstrates the nature of the matching procedure we have used to infer $f_\mathrm{ov}$.
It also highlights that the penetrative convective overshoot, if present, is by far the dominant process near the core, and it is only further from the core that the meridional circulation becomes more important.
However for the purposes of material mixing even the much slower meridional circulation is sufficient to chemically homogenise the system on evolutionary time-scales and so either could be responsible for the inferred chemical mixing near the tachocline.

\section{Implementation in Stellar Models}

While we have not yet implemented these enhanced circulation currents in stellar models, there are a variety of ways in which this could be done.
As a rough approximation one could set the convective overshoot parameters according to equations~\eqref{eq:delta} and~\eqref{eq:alpha1}.
This is only an approximation because the circulation currents exhibit a different spatial dependence from that of convective overshooting, but it is straightforward to do.

A more complete accounting of this mechanism would require tracking the parameter $\alpha'$ in convection zones as well as $\alpha$ in radiative zones.
This may be done either by directly computing these parameters according to the prescriptions in section~\ref{sec:turb} or by expanding the basic equations of stellar structure to incorporate the first several spherical harmonics.
Of particular importance are the harmonics in the heat flux $\boldsymbol{F}$ as well as those in density.
In addition a turbulence closure model is needed to calculate $\D$ in the convection zone, because this mechanism ultimately relies on convective anisotropy.

\section{Interpretation}
\label{sec:interp}

There are two further matters which are worth emphasising.
First, none of what we argue here is in contradiction of the Eddington--Sweet circulation calculations: we are simply extending them to account for a baroclinic boundary condition.
Secondly, the origin of the velocity field we posit lies in turbulence and so the objection that an inviscid fluid cannot support such circulation currents~\citep[see e.g.][]{1981GApFD..17..215B} is simply not applicable.
As we have shown in Appendix~\ref{appen:diffrot} this objection is not even relevant in radiative zones.
Thus we arrive at the conclusion that such circulations are physical.

However there may be magnetic barriers to mixing.
There is strong differential rotation near the solar tachocline so it seems likely that there are also strong toroidal magnetic fields present in this region.
These fields may work to inhibit transverse motion.
At the same time though they serve to increase the turbulent anisotropy, which may partially act to counter this inhibiting effect.
The net effect of magnetic fields in determining the solar meridional flow cannot be too significant because the predicted meridional velocities match observations even in the tachocline.
This suggests that our non-magnetic scaling analysis may suffice, though we cannot say for certain that magnetic fields are generally irrelevant.

\section{Conclusions}

We have demonstrated that the meridional flow rate in the radiative zones of massive rotating stars in many cases is set by anisotropy in the central convection zone.
This enhances mixing in massive stars as well as core-envelope angular momentum coupling in all stars with a radiative zone overlying a convection zone.
Importantly, the additional mixing is approximately independent of the rotation rate above a relatively low threshold, and so ought to be distinguishable from other sources of mixing.

These effects may aid the formation of massive binary black hole pairs~\citep{refId0} through enhanced mixing at long periods.
More generally, such mixing would lead to stronger core-envelope mixing thus leading to massive stars growing larger burnt-out cores and living longer owing to mixing of fuel into the core.
This mechanism mimics convective overshooting, and we have argued that it suffices to explain the anomalous overshoot distances which have been inferred from observations.
In addition, such baroclinicity-driven mixing is likely to occur in stars with outer convection zones and inner radiative zones, though with more limited effect.
This represents a fundamentally new phenomenon in stellar mixing and we expect it to have wide-reaching consequences for stellar evolution.

\section*{Acknowledgements}

The authors gratefully acknowledge many pleasant and productive conversations with Pierre Lesaffre, as well as helpful suggestions from Sterl Phinney, Jim Fuller, Douglas Gough and Steven Balbus.
ASJ acknowledges financial support from a Marshall Scholarship as well as support from the IOA and ENS to work at ENS Paris and from the IOA and CEBS to work at CEBS in Mumbai.
CAT thanks Churchill College for his fellowship.
SMC is grateful to the IOA for support and hopsitality and thanks the Cambridge-Hamied exchange program for financial support.

%%%%%%%%%%%%%%%%%%%%%%%%%%%%%%%%%%%%%%%%%%%%%%%%%%



\bibliographystyle{mnras}
\bibliography{mer.bib} % if your bibtex file is called example.bib

%%%%%%%%%%%%%%%%%%%%%%%%%%%%%%%%%%%%%%%%%%%%%%%%%%

%%%%%%%%%%%%%%%%% APPENDICES %%%%%%%%%%%%%%%%%%%%%

\appendix

\section{Role of Differential Rotation}
\label{appen:diffrot}

The derivation of the Eddington--Sweet circulation relies on it being impossible to satisfy equation\ \eqref{eq:nowindeq} when the system is perfectly barotropic.
One might think from the structure of the argument that this is just a mathematical difficulty rather than a physical one.
After all, valid solutions could exist if one introduces an arbitrarily small differential rotation, allowing the cross-product in equation~\eqref{eq:baro} to be non-zero.

While this is mathematically a valid concern, it runs up against physical difficulties because what matters is how rapidly the diffusivity and other thermodynamic variables may vary along an isobar.
Suppose that equation~\eqref{eq:nowindeq} fails to be satisfied by a dimensionless amount of order $\epsilon$.
Differential rotation may arise and introduce baroclinicity, or misalignment of isobars and isotherms.
We define the angle of misalignment to be $\lambda$~\footnote{See Appendix~\ref{appen:lambda}.}.
If $\lambda \ll 1$, logarithmic derivatives of the diffusivity along isobars must be at least of order $\epsilon/\lambda$ in order to satisfy equation\ \eqref{eq:nowindeq}.
These derivatives ought to be of order unity so we must have $\lambda \approx \epsilon$ or $\lambda > \epsilon$.

To break the degeneracy between differential rotation and meridional circulation we invoke the non-magnetised vorticity equation, which states that
\begin{align}
	\frac{\partial \boldsymbol{\omega}}{\partial t} = \boldsymbol{\omega}\cdot\nabla\boldsymbol{\varv} - \boldsymbol{\omega}\nabla\cdot\boldsymbol{\varv} - \boldsymbol{\varv}\cdot\nabla\boldsymbol{\omega} + \frac{1}{\rho^2}\nabla\rho\times\nabla p + \nabla\times\left(\frac{1}{\rho}\nabla\cdot\mathbfss{T}\right),
	\label{eq:vort0}
\end{align}
where $\boldsymbol{\omega}$ is the vorticity, $\boldsymbol{\varv}$ is the total velocity including rotation, $\rho$ is the density, $p$ is the pressure and $\mathbfss{T}$ is the total fluid stress excluding magnetic effects.
We now let $\boldsymbol{u}$ be the meridional velocity and assume axisymmetry, such that
\begin{align}
	\boldsymbol{\varv} = \boldsymbol{u}(R,z) + \boldsymbol{e}_\phi \Omega(R,z) R,
\end{align}
where $z$ is the cylindrical vertical coordinate and $R$ is the cylindrical radial coordinate.
Then
\begin{align}
	\boldsymbol{\omega}\cdot\nabla\boldsymbol{\varv}
	&= \boldsymbol{\omega}\cdot\nabla\boldsymbol{u} + \boldsymbol{\omega}\cdot\nabla(\Omega R\boldsymbol{e}_\phi)\\
	&= \boldsymbol{\omega}\cdot\nabla\boldsymbol{u} - \boldsymbol{e}_R \Omega \omega_\phi + \boldsymbol{e}_\phi \boldsymbol{\omega}\cdot\nabla (\Omega R),\\
	\boldsymbol{\varv}\cdot\nabla\boldsymbol{\omega}
	&= \boldsymbol{u}\cdot\nabla\boldsymbol{\omega} + \Omega R \boldsymbol{e}_\phi\cdot\nabla\boldsymbol{\omega} \\
	&= \boldsymbol{u}\cdot\nabla\boldsymbol{\omega} - \Omega \omega_\phi \boldsymbol{e}_R + \Omega \omega_R \boldsymbol{e}_\phi,\\
	\intertext{and}
	\boldsymbol{\omega}\nabla\cdot\boldsymbol{\varv}	
	&= \boldsymbol{\omega}\nabla\cdot\boldsymbol{u} + 	\boldsymbol{\omega}\nabla\cdot\left(\boldsymbol{e}_\phi \Omega R\right)\\
	&= \boldsymbol{\omega}\nabla\cdot\boldsymbol{u}.
\end{align}
Putting this together we find
\begin{align}
	&\boldsymbol{\omega}\cdot\nabla\boldsymbol{\varv} - \boldsymbol{\omega}\nabla\cdot\boldsymbol{\varv} - \boldsymbol{\varv}\cdot\nabla\boldsymbol{\omega}\\
	&= \boldsymbol{\omega}\cdot\nabla\boldsymbol{u} - \boldsymbol{u}\cdot\nabla\boldsymbol{\omega} - \boldsymbol{\omega}\nabla\cdot\boldsymbol{u}\\
	&+\boldsymbol{e}_\phi \boldsymbol{\omega}\cdot\nabla(\Omega R)-\Omega \boldsymbol{e}_\phi\omega_R.
\end{align}
With this equation~\eqref{eq:vort0} becomes
\begin{align}
\frac{\partial \boldsymbol{\omega}}{\partial t} = &\boldsymbol{\omega}\cdot\nabla\boldsymbol{u} - \boldsymbol{\omega}\nabla\cdot\boldsymbol{u} - \boldsymbol{u}\cdot\nabla\boldsymbol{\omega}\nonumber\\
& + \boldsymbol{e}_\phi \boldsymbol{\omega}\cdot\nabla(\Omega R)-\Omega \boldsymbol{e}_\phi\omega_R\nonumber\\
& + \frac{1}{\rho^2}\nabla\rho\times\nabla p + \nabla\times\left(\frac{1}{\rho}\nabla\cdot\mathbfss{T}\right).
\label{eq:vorticity1}
\end{align}
Suppose that differential rotation which satisfies equation~\eqref{eq:nowindeq} arises with no meridional flow.
So $\boldsymbol{u}=0$ and
\begin{equation}
	\boldsymbol{\omega} = \boldsymbol{e}_z \frac{1}{R}\frac{\partial(\Omega R^2)}{\partial R} - \boldsymbol{e}_R R \frac{\partial \Omega}{\partial z}.
\end{equation}
This rotation must then also satisfy equation~\eqref{eq:vorticity1}.
In steady state this reduces to
\begin{align}
	0 = \boldsymbol{e}_{\phi}\boldsymbol{e}_{z}\cdot\nabla (\Omega^2 R) + \frac{1}{\rho^2}\nabla\rho\times\nabla p + \nabla\times\left(\frac{\nabla\cdot\mathbfss{T}}{\rho}\right).
	\label{eq:vort}
\end{align}
We refer to the first term as advective because it is related to the advection of angular momentum.
The second is the thermal wind term~\citep{2012MNRAS.420.2457B} and the third reflects the turbulent stress.

Because equation~\eqref{eq:vort} is a vector equation it contains three scalar equations,
\begin{align}
	\label{eq:diffrot}
	0 &= \boldsymbol{e}_{z}\cdot\nabla (\Omega^2 R) + \frac{1}{r \rho^2}\left(\frac{\partial \rho}{\partial r}\frac{\partial P}{\partial \theta}-\frac{\partial P}{\partial r}\frac{\partial \rho}{\partial \theta}\right)+ \boldsymbol{e}_\phi\cdot\nabla\times\left(\frac{\nabla\cdot\mathbfss{T}}{\rho}\right),\\
	\label{eq:T1}
	0 &= \boldsymbol{e}_R\cdot\nabla\times\left(\frac{\nabla\cdot\mathbfss{T}}{\rho}\right)\\
\intertext{and}
	\label{eq:T2}
	0 &= \boldsymbol{e}_z\cdot \nabla\times\left(\frac{\nabla\cdot\mathbfss{T}}{\rho}\right),
\end{align}
where $r$ is the spherical radial coordinate.
Given $P$ and $\rho$ satisfying equation\ \eqref{eq:nowindeq} it is possible to solve for $\Omega$ with the first of these equations because $\boldsymbol{\omega}$ is just a function of $\Omega$.
The remaining two equations must then be satisfied by that solution.
Because they depend on the turbulent flux there is no reason to expect them to be satisfied by a rotation profile arrived at independently.
As a result the vorticity equation is unlikely to be satisfied without a meridional flow.

From this we conclude that satisfying both the vorticity equation and thermal equilibrium requires a meridional flow.
Now suppose that differential rotation permits baroclinicity of order $\lambda$.
If $\lambda \ll \epsilon$ then thermal equilibrium relies on the meridional flow.
If $\lambda \gg \epsilon$ then thermal equilibrium could be satisfied without a meridional flow.
Because the flow appears in the vorticity equation at first order multiplied by the rotation, and the differential rotation also appears at first order multiplied by the rotation, the flow must be at least as large as the differential rotation, possibly larger if turbulent stresses dominate.
In this case then the flow is at least of order $\lambda$ and hence also suffices to satisfy thermal equilibrium.
This means that, no matter what, there is a meridional flow which is at least of order $\epsilon$.

In the remainder of this appendix we show that if thermal equilibrium is satisfied without a meridional flow the differential rotation and/or baroclinicity which results {\em mechanically} drives a meridional flow.
We furthermore show that this flow exceeds in magnitude that which would have been required to satisfy thermal equilibrium, and hence conclude that a flow of this order ought to generally arise.

\subsection{Radiative Zones}

We first consider radiative zones.
To begin note that because the system is axisymmetric the thermal wind term has magnitude
\begin{align}
	\frac{1}{\rho^2} \left|\boldsymbol{e}_\phi \cdot \nabla p \times \nabla \rho\right| &= \frac{1}{\rho^2}|\nabla p \times \nabla \rho|\\
	&= \frac{\lambda}{\rho^2 \gamma}|\nabla P| |\nabla \rho|,
\end{align}
where we have used the definition $\lambda$ as the small angle between the density and pressure gradients.
Taking both derivatives to produce factors of order $h^{-1}$ we find
\begin{align}
	\frac{1}{\rho^2}|\nabla p \times \nabla \rho| &\approx \frac{\lambda P}{h^2 \rho^2 \gamma}.
\end{align}
Using $P = \rho g h$ and neglecting $\gamma$ because it is of order unity yields
\begin{align}
	\left|\boldsymbol{e}_\phi \cdot (\nabla p \times \nabla \rho)\right| &\approx \lambda \frac{g}{h}.
\end{align}

There are now two cases to consider.
First suppose that in radiative zones the stress term in equation~\eqref{eq:diffrot} balances the thermal wind term.
Then
\begin{align}
	\left|\boldsymbol{e}_\phi\cdot\nabla\times\left(\frac{\nabla\cdot\mathbfss{T}}{\rho}\right)\right| \approx \frac{\mathsf{T}}{\rho h^2} \approx \frac{g}{h}\lambda,
	\label{eq:stress0}
\end{align}
where we have approximated derivatives by factors of $h^{-1}$ and where $\mathsf{T} = |\mathbfss{T}| = \max(\mathsf{T}_{ij})$.
Using equation~\eqref{eq:vort0} we may now calculate the circulation implied by the vorticity equation.
Because the meridional circulation enters the meridional components of the vorticity equation in the form of gradients multiplied by $\Omega$, we have
\begin{align}
	\label{eq:nou}
	u &\approx \frac{1}{h\rho\Omega}\mathsf{T}\\
	\label{eq:u}
	&\approx \lambda\frac{g}{\Omega}.
\end{align}
In essence we are balancing terms like $\boldsymbol{\omega}\cdot\nabla\boldsymbol{u}$ and $\boldsymbol{\omega}\nabla\cdot\boldsymbol{u}$ with the stress induced by the differential rotation and using $|\boldsymbol{\omega}| \approx \Omega$.
The rate at which heat is deposited per unit volume by this flow is of order 
\begin{align}
	Q = \rho c_p T u |\nabla s| \approx \rho c_p T \frac{u}{h},
\label{eq:QQQ}
\end{align}
because, in radiative zones, the dimensionless entropy gradient is on the order of $h^{-1}$.
Noting that $c_p$ is of order $k_{\rm B}$ and so $P \approx \rho c_p T$ for an ideal gas, we find
\begin{align}
	Q \approx \lambda P \frac{g}{h \Omega}.
\end{align}
The flux anisotropy is
\begin{align}
	\delta F \approx \epsilon F.
\end{align}
This deposits heat at a rate
\begin{align}
	\delta Q = \nabla\cdot \delta\boldsymbol{F} \approx \epsilon \frac{F}{h},
\end{align}
where we have used the fact that the characteristic scale of the flux anisotropy is the pressure scale height.
So
\begin{align}
	\frac{Q}{\delta Q} \approx \frac{\lambda}{\epsilon}\left(\frac{g P}{\Omega F}\right) > \frac{g P}{\Omega F} = \frac{P\sqrt{gh}}{F}\left(\frac{g}{\Omega^2 h}\right)^{1/2} \gg \frac{P}{F}\sqrt{gh},
\end{align}
when $\lambda \geq \epsilon$ and $\Omega^2 \ll g/h$.
Noting that
\begin{align}
	\sqrt{gh} = \sqrt{\frac{P}{\rho}}= c_s \sqrt{\gamma},
\end{align}
where $c_s$ is the adiabatic sound speed, and that the heat flux in a radiative zone is bounded above\footnote{To see this note that $F = 16\sigma |\nabla T^4|/3\kappa \rho$. If the medium is optically thick over a scale height then $\kappa \rho h > 1$, so $F < 16 \sigma h|\nabla T^4| = 4 a c T^4 h |\nabla \ln T|$. The onset of convection occurs when $|\nabla \ln T| > (2/5) |\nabla \ln P| = 2/5h$, so $F < 8 a c T^4/5$. Noting that radiation pressure is given by $P_{\rm rad}=aT^4/3$, $F/P < (24/5)c (P_\mathrm{rad}/P)$. The sound speed is $c_\mathrm{s} = \sqrt{\gamma P/\rho} = c \sqrt{\gamma P/\rho c^2}$, so $F/P c_\mathrm{s} \la (P_\mathrm{rad}/P)\sqrt{P/\rho c^2}$. The first factor is at most unity by definition and the second is much less than unity (for non-relativistic gases) or of order unity (for relativistic gases), so the bound holds.} by $c_s P$, we find that $Q > \delta Q$.
The heat transported in this way is therefore greater than the initial flux disturbance.
As a result this flow does more to relieve the violation of thermal equilibrium than the differential rotation and so, in practice, a much milder differential rotation develops, with the circulation on the order of the baroclinicity.

A similar outcome can be seen in the case where the advective term balances the thermal wind term.
Recall that the advective term is
\begin{align}
	\boldsymbol{e}_z \cdot \nabla (\Omega^2 R) \approx \Omega^2 |R\nabla\ln\Omega|,
\end{align}
where we have assumed that the differential rotation is not strongly preferentially along $\boldsymbol{e}_R$.
In this case
\begin{align}
	|R\nabla\ln \Omega| \approx \lambda \frac{g}{h\Omega^2}.
	\label{eq:drt}
\end{align}
Noting that $\lambda \geq \Omega^2 r/g$ we find\footnote{We show in Appendix~\ref{appen:centrifugal} that $\lambda$ receives a contribution from the centrifugal term of order $\Omega^2 r/g$, but there may be additional contributions from other effects such as flux anisotropies. For the purposes of this argument though it suffices to have a lower bound.}
\begin{align}
	|R\nabla\ln \Omega| \geq \frac{r}{h} \approx 1,
	\label{eq:drt1}
\end{align}
where we have approximated $r/h$ as unity because this is the case in the bulk of the star.
Now note that the turbulent viscosity is of order~\citep{1997A&A...321..134M}
\begin{equation}
	\nu \approx \xi \chi\frac{S^2}{N^2},
	\label{eq:nu}
\end{equation}
where $\xi$ is a dimensionless constant of order unity, $S$ is the shear, $N^2 \approx g/h$ is the square of the \brvs\ frequency and
\begin{equation}
	\chi = \frac{k}{\rho c_p}
	\label{eq:chi}
\end{equation}
is the thermal diffusivity with $k$ the thermal conductivity.
We are able to use this expansion because the differential rotation which balances the thermal wind term is vertical and so not subject to centrifugal stabilisation.
In order for this prescription to apply however we require that
\begin{align}
	\frac{N^2}{c' S^2}\left(\frac{\nu_{\rm micro}}{\chi}\right) < 1,
	\label{eq:crit}
\end{align}
where $\nu_{\rm micro}$ is the microscopic viscosity and $c'$ is a constant which has been variously estimated as lying between $10^{-2}$ and $1$~\shortcite{2018arXiv180311530K}.
We shall later verify that this holds in most cases of interest.
Even when it does not a comparable diffusivity may be provided by turbulence in the non-stratified direction~\shortcite{1993ssrv...66..285Z}.
Following equation (36) of~\citet{2004ApJ...607..564M} we find that so long as the colatitude $\theta$ is such that $\sin^2\theta$ is small compared with the reciprocal of the left-hand side of equation~\eqref{eq:crit} this effect is active.
Furthermore it is active if the specific angular momentum decreases outward.
Hence even when equation~\eqref{eq:crit} is not precisely satisfied equation~\eqref{eq:nu} likely still applies in much of the star.

Taking $\xi=1$ and noting that
\begin{equation}
	\mathsf{T} \approx \rho \nu S
\end{equation}
we find
\begin{equation}
	\rho^{-1} \mathsf{T} \approx \frac{k h}{\rho c_p g} S^3.
\end{equation}
In a system with differential rotation
\begin{equation}
	S \approx |R\nabla\Omega|
		\label{eq:Sestim}
\end{equation}
so
\begin{align}
	\rho^{-1}\mathsf{T} \approx \frac{k h}{\rho c_p g} \Omega^3 |R\nabla\ln \Omega|^3.
\end{align}
Using equations~\eqref{eq:drt} and the first equality of~\eqref{eq:u} we may now calculate the circulation implied by the vorticity equation.
So we have
\begin{align}
	u \approx \frac{1}{h\rho\Omega}\mathsf{T} \approx \lambda^3 \frac{k}{\rho c_p g} \frac{g^3}{h^3\Omega^4}.
\end{align}
Noting that $F = |k\nabla T|$ and $|\nabla T| \approx T/h$ we find
\begin{align}
	u \approx \lambda^3 \frac{F}{\rho c_p T} \frac{g^2}{h^2\Omega^4},
	\label{eq:uest}
\end{align}
Noting again that $\lambda \ga \Omega^2 r/g$,
\begin{align}
	u \ga \lambda \frac{F}{\rho c_p T} \frac{g}{h\Omega^2}\left(\frac{r}{h}\right)^2.
\end{align}
Because the star is spinning below breakup, $\sqrt{g/h}$ is greater than $\Omega$.
We also have $r > h$ so, when $\lambda \geq \epsilon$,
\begin{align}
	u &>  \lambda \frac{F}{\rho c_p T} \geq \epsilon \frac{F}{\rho c_p T}.
	\label{eq:ubound}
\end{align}
The heat deposited by this flow is again~\eqref{eq:QQQ} of order 
\begin{align}
	Q = \rho c_p T u |\nabla s| \approx \rho c_p T \frac{u}{h},
\end{align}
because, in radiative zones, the dimensionless entropy gradient is on the order of $h^{-1}$.
With equation~\eqref{eq:ubound} we find
\begin{align}
	Q > \epsilon \frac{F}{h} \approx \delta Q.
\end{align}
Once more the heat transported by the circulation is greater than the initial flux disturbance, so we expect a much milder differential rotation to develop, with the circulation on the order of the baroclinicity.

We now determine the conditions under which inequality~\eqref{eq:crit} gives rise to a vertical instability.
Using equation~\eqref{eq:Sestim} this may be written as
\begin{align}
	\frac{N^2}{|R\nabla\Omega|^2}\left(\frac{\nu_{\rm micro}}{\chi}\right) < c'.
\end{align}
Near the cores of stars, which is where we shall ultimately be interested in this expression, the viscosity is predominantly radiative\footnote{Higher up in the atmosphere the particle viscosity dominates, but there $h/r$ is small and equation~\eqref{eq:drt1} indicates that the shear is large, so we do not expect that limit to violate equation~\eqref{eq:crit}.}, so that~\shortcite{doi:10.1093/mnras/86.5.328}
\begin{align}
	\nu_{\rm micro} \approx \frac{\chi c_{\rm p} T}{4 c^2},
\end{align}
where $c \approx 3\times 10^{10} \mathrm{cm\,s^{-1}}$ is the speed of light.
Hence the condition is that
\begin{align}
	\frac{N^2}{|R\nabla\Omega|^2}\left(\frac{c_{\rm p} T}{4 c^2}\right) < c'.
\end{align}
We now need a better estimate of the differential rotation than that provided by equation~\eqref{eq:drt1}.
Near the cores of these stars, where equation~\eqref{eq:crit} is hardest to satisfy because $N^2$ and $T$ are largest, we argue in Section~\ref{sec:turb} that $\epsilon \approx \Omega^2/|N|_{\rm core}^2$, with an upper bound of order unity, where $|N|^2_{\rm core}$ is the average \brvs\ frequency in the convection zone.
Hence if $\lambda \geq \epsilon$ then with equation~\eqref{eq:drt} we find
\begin{align}
	\frac{\min(|N|_{\rm core}/\Omega,1)^2 N^2 h}{ g}\left(\frac{c_{\rm p} T}{4 c^2}\right) < c'.
\end{align}
Noting that $g/h \approx N^2$ we find
\begin{align}
	\min\left(1,\frac{|N|_{\rm core}^2}{\Omega^2}\right)\left(\frac{c_{\rm p} T}{4 c^2}\right) < c'.
\end{align}
Noting that $c_p T \approx c_{\rm s}^2$, we obtain
\begin{align}
	\min\left(1,\frac{|N|_{\rm core}^2}{\Omega^2}\right)\left(\frac{c_{\rm s}^2}{4 c^2}\right) < c',
\end{align}
or
\begin{align}
	\min\left(1,\frac{|N|_{\rm core}}{\Omega}\right)\left(\frac{c_{\rm s}}{c}\right) < 2c',
\end{align}
In the cores of massive stars such as those we consider in Section~\ref{sec:massive}, $c_{\rm s} \approx 3^{7}\mathrm{cm^2\,s^{-2}}$ and we take $R \approx 0.2 R_\odot \approx 10^{10}\mathrm{cm}$ so
\begin{align}
	\min\left(1,\frac{|N|_{\rm core}}{\Omega}\right) < 2\times 10^3 c'.
\end{align}
Using $2\times 10^3 c' > 1$ this reduces to
\begin{align}
	\Omega > \frac{|N|_{\rm core}}{2\times 10^3 c'}.
\end{align}
In the stars of interest $|N|_{\rm core} \approx 10^{-8}{\rm s^{-1}}$, so even with the pessimistic estimate $c' \approx 10^{-2}$ the cutoff frequency is around $10^{-9}{\rm s^{-1}}$, which accommodates the vast majority of stars.
Away from the core we show in Section~\ref{sec:massive} that the flux anisotropy drops off so the shear falls to that given by equation~\eqref{eq:drt1}.
As this occurs $h$ falls, equation~\eqref{eq:drt1} becomes a tighter bound, $N T$ falls and the criterion becomes looser, so we do not expect this to represent a significant limitation, though there could be stars for which equation~\eqref{eq:crit} fails at some intermediate point.
Such objects are potentially of interest, though even when the criterion is not precisely satisfied horizontal turbulence of comparable magnitude may still be active in a region in the vicinity of the poles.
This means that at a minimum turbulence is present in many stars, even those which rotate quite slowly, though the latitudinal extent of the circulation may be restricted in the slowest-rotating stars.

A similar and in many ways more detailed analysis was made by~\citet{doi:10.1093/mnras/stv115}, but they neglected turbulent stresses, which we find to be crucial to the angular momentum balance.
~\citet{2016MNRAS.457.1711C} studied linear modes in the solar radiative zone and found them to be asymptotically stable, though many such modes grow significantly on the short term. This is consistent with the findings of~\citet{ASL:ASL153} who show that such modes which are not asymptotically unstable nevertheless generate turbulence for arbitrarily small shears, scaling in the manner predicted by~\citet{1986A&A...168...89C} and~\citet{1993ssrv...66..285Z}. This has been verified in simulations~\citep{pratetal} and is what we assume here.

Likewise, it is useful to note that this argument may be recast in the language of~\citet{2010ApJ...719..313G} by writing their parameter
\begin{align}
	\sigma \equiv \frac{N}{\Omega}\sqrt{\frac{\nu}{K}} = \frac{S}{\Omega}\sqrt{\xi} \approx \frac{S}{\Omega}.
\end{align}
It follows that equation~\eqref{eq:relshear} places an upper bound on $\sigma$ of order $S/\Omega \approx |R\nabla\ln \Omega|$.
So long as this is at least of order unity the flow is not mechanically constrained~\citep{doi:10.1046/j.1365-8711.2002.05662.x}, and we have argued that this is the case.
The scale over which mechanically forced perturbations decay is of order $R/\sigma$~\citep{2010ApJ...719..313G}, so should $\sigma$ be large this would be quite small.
We shall argue below that $\sigma$ is of order unity but may in principle grow large, in which case this may set constraints on the extent of the circulation.
However because the flow in this case is locally forced by turbulent stresses it is less clear if this scale applies.
In particular, because the stresses are driven by differential rotation and baroclinicity, which in turn result from perturbations to the heat flux, the scale on which perturbations mechanically damp is much less important than that on which the heat flux becomes spherically symmetric.
We find this to be of order $r$ in Section~\ref{sec:massive} and which suggests that $\sigma$ does not actually set the decay length for this circulation.

Finally, with knowledge of the nature of turbulence in these systems we can, in principle, infer the extent of the differential rotation which arises.
In practice our knowledge of these instabilities is incomplete, and we have entirely neglected magnetic fields, so the result of such an inference may not be trustworthy.
Nevertheless it is interesting to consider and we do so now.
Because baroclinic instabilities are unlikely to be active in these systems~\citep{1984A&A...132...89S} it seems more likely that the azimuthal vorticity balance is between the advective terms balance the baroclinic one and that the resulting differential rotation generates shear turbulence.
In this case the baroclinicity is reduced by the meridional circulation such that the shears drive a circulation which is just able to ensure thermal equilibrium.
Setting $Q = \delta Q$, we see from equation~\eqref{eq:uest} that
\begin{align}
	\lambda^3 \frac{g^2}{h^2\Omega^4} \approx \epsilon,
\end{align}
so that all subsequent inequalities become equalities.
Hence the baroclinicity is of order
\begin{align}
	\lambda \approx \left(\frac{\epsilon h^2 \Omega^4}{g^2}\right)^{1/3},
\end{align}
which is quite small even for $\epsilon$ of order unity.
With this we also find
\begin{align}
	\frac{S}{\Omega} \approx |R\nabla\ln\Omega| \approx \left(\frac{\epsilon g}{h\Omega^2}\right)^{1/3},
	\label{eq:relshear}
\end{align}
so as $\Omega \rightarrow 0$ the differential rotation becomes small but the relative differential rotation becomes large.
This behaviour cuts off when $\Omega \approx |N|$ in the convection zone, at which point $\epsilon \approx \epsilon_{\rm max}$ and
\begin{align}
	|R\nabla\ln\Omega| \approx \left(\frac{g \epsilon_{\rm max}}{h |N|_{\rm c, core}^2}\right)^{1/3} \approx \left(\epsilon_{\rm max}\frac{c_{\rm s}^2}{u_{c,{\rm core}}^2}\right)^{1/3}.
\end{align}
For the stars of interest $c_{\rm s}$ in the core is of order $3\times 10^7{\rm cm\,s^{-1}}$ and $u_{c,{\rm core}} \approx 10^{3}{\rm cm\,s^{-1}}$.
Taking $\epsilon_{\rm max} \approx 0.2$ and we find the limiting relative differential rotation is $|R\nabla\ln\Omega| \la 300$, corresponding to relatively mild absolute shears of order $3\times 10^{-6}{\rm s^{-1}}$.
This represents an upper bound  on the {\emph relative shear} in that any additional stresses or instabilities would serve to reduce the requisite shear.
For the more typical case of a star with $h \approx R/2 \approx R_\odot$ rotating at $30{\rm km\,s^{-1}}$ equation~\eqref{eq:relshear} gives $|R\nabla\ln\Omega| \approx 3$.

In practice we do not expect the differential rotation to be so extreme.
Other factors, notably the generation of a magnetic field by dynamo action, ought to become increasingly important as the rotation rate decreases and serve to further bound the differential rotation~\citep{2002A&A...381..923S,2005A&A...440L...9E}.
It is also possible that when the differential rotation becomes large it serves to in some fashion suppress the convective flux anisotropy by providing mechanical feedback to the convection zone.
Without a robust formalism for evaluating these possibilities, however, we leave them for the future.

\subsection{Convection Zones}

A similar argument holds in convection zones with two modifications.
First, without any differential rotation or circulation currents, the meridional components of the vorticity equation are imbalanced due to the $\Lambda$-effect~\citep{2013IAUS..294..399K}.
We neglect this effect because it does not generically balance the differential rotation which emerges from equation~\eqref{eq:diffrot} because that has a physically distinct origin and hence, in either case, there is a conflict between the meridional and azimuthal components of the vorticity equation which must be resolved by a meridional circulation.
When, as we assume, there is no geometric tuning between the thermal wind term and the $\Lambda$-effect the magnitude of this conflict is set by the greater of the two effects, and hence must be at least that implied by equation~\eqref{eq:diffrot}.
Thus we may reasonably find a lower bound on the circulation velocity implied by the requirements of thermal equilibrium without considering the $\Lambda$-effect.

The second modification is to account for the fact that the flux in convection zones is modulated by the entropy gradient rather than by the temperature gradient.
The entropy gradient is reduced relative to the bare thermodynamic gradients by a factor of $|N|^2 h/g$, and so the baroclinicity required to restore thermal equilibrium is reduced by the same factor.
Thus $\lambda \geq \epsilon |N|^2 h/g$.

With these modifications in hand we may begin.
Once more there are two cases.
First suppose that $\Omega < |N|$.
Proceeding as before we note that in convecting regions the differential rotation perturbs the stresses by an amount of order $\nu |R\nabla\Omega|$, where $\nu \approx h^2 |N|$~\citep{gough78} is the turbulent viscosity.
Thus
\begin{align}
	\rho^{-1}\mathsf{T} &\approx \nu |\nabla \varv|\\
	&\approx \nu |R\nabla\Omega|\\
	&\approx \nu \Omega|R\nabla\ln\Omega|\\
	&\approx h^2 \Omega  |N| |R\nabla\ln \Omega|.
	\label{eq:TA}
\end{align}
Because $|N| > \Omega$ this term, rather than the advective one, balances the thermal wind term in equation~\eqref{eq:diffrot}.
Thus
\begin{align}
	\rho^{-1}\mathsf{T} &\approx \lambda h g.
\end{align}
From equation~\eqref{eq:u} we have
\begin{align}
	u &\approx \frac{1}{h\rho\Omega}\mathsf{T}\\
	&\approx \lambda\frac{g}{\Omega}\\
	&\approx g \frac{|N|}{\Omega |N|} \lambda\\
	&\approx h  |N| \frac{g/h}{\Omega |N|} \lambda,
\end{align}
where $\varv_\mathrm{c}$ is the convection speed.
Because $g/h \gg \Omega^2$ and $|N| > \Omega$ we have
\begin{align}
	u&\approx \varv_\mathrm{c} \frac{g/h}{\Omega |N|} \lambda\\
	&\geq \epsilon u_\mathrm{c} \frac{|N|}{\Omega}\\
	&\geq \epsilon u_\mathrm{c}.
\end{align}
Because $\varv_\mathrm{c}$ sets the scale of heat transport it follows that $u$ more than suffices to transport the flux anisotropy $\epsilon$.

Next suppose that $\Omega > |N|$.
In this case the thermal wind term $\lambda g/h$ is of order $|N|^2$, because the flux anisotropy cannot be larger than of order unity~\citep{jermyn}.
If we balance this with the advective term we find
\begin{align}
	|R\nabla\ln\Omega| \approx \frac{|N|^2}{\Omega^2},
\end{align}
while if we balance it with the turbulent stress we find with equation~\eqref{eq:TA}
\begin{align}
	|R\nabla\ln\Omega| \approx \frac{|N|}{\Omega}.
\end{align}
The former is smaller because the advective term is larger and so it is the advective term which accomplishes this balance.
The stress is therefore
\begin{align}
	\rho^{-1}\mathsf{T}
	&\approx h^2 |N| \Omega |R\nabla\ln \Omega|\\
	&\approx h^2 |N|^2.
\end{align}
So, as when $\Omega < |N|$,
\begin{align}
	u &\approx \frac{1}{h\rho\Omega}\mathsf{T}\\
	&\approx h |N|\frac{|N|}{\Omega}\\
	&\approx u_\mathrm{c} \frac{|N|}{\Omega}.
\end{align}
In this case we find instead that the vorticity equation can be balanced by a circulation which does not balance the flux anisotropy.
Thus in the case of rapid rotation baroclinicity may play a more significant role than the circulation in balancing the heat equation.

It follows then that in all but the case of rapidly rotating convection the dominant effect in restoring thermal equilibrium is the circulation current and not the introduction of baroclinicity and differential rotation.
The restoring circulation therefore takes on the characteristic scale of the violation of thermal equilibrium, in this case scaling like the velocity associated with heat transport and $\epsilon$.

It is instructive to compare this conclusion with the results of three-dimensional simulations of rotating convection zones.
These simulations typically find a circulation current somewhat larger than that required to maintain thermal equilibrium yet of the same order~\citep{doi:10.1146/annurev.fluid.010908.165215}.
This has been attributed to the requirements of the vorticity equation in the presence of anisotropic turbulent stresses\footnote{These include the $\Lambda$-effect.} because, though such effects scale in the same way as the flux ansiotropy, they are generally found to be several times larger.
However it is also possible that the flux anisotropy is ultimately the driver of these circulations.
Nothing in our arguments precludes a larger velocity than the minimum which thermal imbalance requires, so an alternate interpretation of the simulations is that the requirement of thermal equilibrium pumps energy into the circulation and the flow geometry then arranges itself to match this input to the dissipation.
The final circulation rate is not constrained except insofar as it must be at least the minimum required to restore thermal equilibrium\footnote{Otherwise the energy input would rise as a thermal imbalance developed.}.

\section{Baroclinicity}
\label{appen:lambda}

In this Appendix we consider the contribution to the baroclinicity owing to both convective anisotropy and centrifugal effects.
First though we define the angle $\lambda$ between the pressure and entropy gradients as
\begin{align}
	\lambda \equiv \sin^{-1}\left(\frac{\left|\nabla P \times \nabla \rho\right|}{|\nabla P||\nabla \rho|}\right).
\end{align}
Noting that in a nearly-adiabatic region
\begin{equation}
	|\nabla \ln \rho| \approx \gamma^{-1}|\nabla \ln P|,
\end{equation}
where $\gamma$ is the adiabatic index, we find that for small angles
\begin{align}
	\lambda 
&\approx \frac{\left|\nabla \ln P \times \nabla \ln \rho\right|}{|\nabla \ln P||\nabla \ln \rho|}\\
&\approx \gamma\frac{\left|\nabla \ln P \times \nabla \ln \rho\right|}{|\nabla \ln P|^2}.
\end{align}
Using the fact that $\nabla P$ and $\nabla \rho$ both lie in the meridional plane we find
\begin{align}
\lambda
&\approx \gamma\frac{\left|\boldsymbol{e}_{\phi}\cdot\left(\nabla \ln P \times \nabla \ln\rho\right)\right|}{|\nabla \ln P|^2}\\\
&\approx \gamma \left|\boldsymbol{e}_{\phi}\cdot\left((h \nabla \ln P)\times(h\nabla \ln \rho)\right)\right|,
\end{align}
where
\begin{equation}
	h \equiv |\nabla \ln P|^{-1}
\end{equation}
is the pressure scale height.
If we approximate the pressure gradient as radial then
\begin{align}
\lambda 
&\approx \left|\boldsymbol{e}_{\phi}\cdot\left(-\boldsymbol{e}_{r}\times(h\nabla \ln \rho)\right)\right|\\
&\approx \frac{h}{r}|\partial_\theta \ln \rho|.
\end{align}
For an ideal gas we may write\footnote{In massive stars the effects of radiation pressure may matter, but these just serve to alter $\gamma$ and do not fundamentally change our analysis, so we omit them for simplicity.}
\begin{equation}
	s = \ln \frac{P}{\rho^\gamma},
	\label{eq:s}
\end{equation}
so
\begin{equation}
	\lambda \approx \gamma h |s_\theta|.
	\label{eq:lambda0}
\end{equation}
This parametrises the baroclinicity.
Because it is of order unity we neglect the factor of $\gamma$ and use equation~\eqref{eq:pert} to find
\begin{align}
	\lambda \approx h |s_r \alpha|.
	\label{eq:lambda1}
\end{align}
This is our leading order estimate of the baroclinicity in the convection zone.

\section{Centrifugal Acceleration}
\label{appen:centrifugal}

The contribution to baroclinicity from the centrifugal acceleration is quite a bit smaller than that from convective anisotropy.
The pressure gradient distorts by an amount of order $\rho \Delta(\nabla \Phi) \approx \rho \Omega^2 R$.
If the density were not to adjust as well this would influence $\lambda$ as
\begin{equation}
	\Delta \lambda \approx h \Delta(\nabla \ln p) \approx \frac{h}{p}\rho \Omega^2 R = \frac{\Omega^2 R}{g}.
\end{equation}
However in reality the density gradient adjusts too and it is instead the entropy gradient which is perturbed by the requirement of thermal equilibrium~\shortcite{1929MNRAS..90...54E}.
The corresponding perturbation to the baroclinicity is therefore multiplied by a factor of $h s_r$, so
\begin{equation}
	\Delta \lambda \approx \frac{\Omega^2 R}{g} h s_r,
\end{equation}
which is much less than $h s_r \alpha$ because $N^2 \ll g/R$.
Thus the total baroclinicity is well approximated by equation~\eqref{eq:lambda1}.

\section{Damping}
\label{appen:damp}

The term that appears in equation~\eqref{eq:thermalBalance4} associated with the meridional flow is
\begin{align}
	Q_\mathrm{flow} &\equiv \frac{r}{D_0 s_r} \left(u_r s_r + u_\theta s_\theta\right)\\
	&=\frac{r}{D_0} \left(u_r + u_\theta \frac{s_\theta}{s_r}\right).
\end{align}
Inserting equations~\eqref{eq:ur0} and~\eqref{eq:ut0} we find
\begin{align}
	Q_\mathrm{flow} \approx \alpha' + \alpha'\frac{r}{h} \frac{s_\theta}{s_r},
\end{align}
and with equation~\eqref{eq:pert} we arrive at
\begin{align}
	Q_\mathrm{flow} \approx \alpha' + \alpha'\alpha\frac{r}{h}.
\end{align}
Because both $\alpha$ and $\alpha'$ are quadratic in the rotation we drop the second term and find
\begin{align}
	Q_\mathrm{flow} \approx \alpha'.
\end{align}

Because the terms in equation~\eqref{eq:thermalBalance4} are those which contribute to $\alpha$ and $\alpha'$, and because this must act to damp the baroclinicity\footnote{This is because the work which powers the flow comes from the baroclinicity, or equivalently the flux anisotropy, and so thermodynamically the only option is for the flow to damp it.}, we find
\begin{align}
	\frac{\mathrm{d}\alpha}{\mathrm{d}\ln P} 
&= Q_\mathrm{flow} - \frac{\mathrm{d}}{\mathrm{d} \ln P}\left[\int_{\ln P}^{\ln P_\mathrm{c}} \frac{\Omega(\mathscr{P})^2}{|N|(\mathscr{P})^2}\mathrm{d}\ln \mathscr{P}\right]\\
&= \alpha' - \frac{\Omega^2}{|N|^2}.
\end{align}
Because the term proportional to $\alpha'$ is responsible for damping $\alpha'$, the coefficient may matter, so we write
\begin{align}
	\frac{\mathrm{d}\alpha}{\mathrm{d}\ln P} &= y' \alpha' - \frac{\Omega^2}{|N|^2},
\end{align}
where $y'$ is a dimensionless factor of order unity determined by the geometry of the circulation.
With equation~\eqref{eq:ap} this becomes
\begin{equation}
	\frac{\mathrm{d}\alpha}{\mathrm{d}\ln P} = y \alpha - \frac{\Omega^2}{|N|^2},
\end{equation}
where $y$ is another dimensionless factor of order unity.
Applying the boundary condition $\alpha(P_\mathrm{c}) = 0$, which must be true because all latitudinal derivatives vanish at the origin, we find
\begin{equation}
	\alpha = P^y\int_{P}^{P_\mathrm{c}} \frac{\Omega(\mathscr{P})^2}{|N|(\mathscr{P})^2} \frac{\mathrm{d}\mathscr{P}}{\mathscr{P}^{1+y}}
\end{equation}
as desired.

One might worry that the turbulent diffusivity found in Appendix~\ref{appen:diffrot} could also contribute significantly to the damping, but this is not the case.
The turbulent thermal diffusivity is of order the viscosity given by equation~\eqref{eq:nu}, so
\begin{align}
	\chi_{\rm turb} \approx \nu
\end{align}
Using equation~\eqref{eq:nou} we find the circulation to be of order
\begin{align}
	u \approx \frac{\mathsf{T}}{h\rho \Omega} \approx \nu\frac{S}{h \Omega},
\end{align}
producing effective thermal diffusivity
\begin{align}
	\chi_{\rm circulation} \approx u_r h \approx \nu \frac{S}{\Omega}\left(\frac{h}{r}\right). 
\end{align}
Near the cores of stars $h \approx r$ and we have argued that $S \ga \Omega$, so this is larger than the turbulent thermal diffusivity, so the latter may safely be neglected.

%%%%%%%%%%%%%%%%%%%%%%%%%%%%%%%%%%%%%%%%%%%%%%%%%%

% Don't change these lines
\bsp	% typesetting comment
\label{lastpage}
\end{document}